\documentclass[12pt,a4paper]{article}
\pdfoutput=1
\usepackage{amsmath}
\usepackage{amssymb}
\usepackage[compress]{cite}
\usepackage[]{graphicx}
\graphicspath{{figures/}}
\usepackage{hyperref}
\usepackage{xcolor}
\usepackage{authblk}

\usepackage{microtype}
\usepackage{booktabs}

\DeclareMathOperator{\im}{Im}
\DeclareMathOperator{\re}{Re}
\DeclareMathOperator{\arsinh}{arsinh}
\DeclareMathOperator{\Li2}{Li_2}
\newcommand{\PFQ}[5]{\ensuremath{
 {}_{#1}F_{#2}\bigg(
 \begin{array}{c}
   \begin{array}{*{#1}c}
     #3
   \end{array}\\
   \begin{array}{*{#2}c}
     #4
   \end{array}
 \end{array}\bigg|#5\bigg)
}}

\numberwithin{equation}{section}

\newcommand{\MS}{\ensuremath{\overline{\text{MS}}}}

\begin{document}

\allowdisplaybreaks
\thispagestyle{empty}

\begin{flushright}
{\small
TUM-HEP-966/14\\
SFB/CPP-14-88\\
arXiv:1411.3132\\
December 16, 2014}
\end{flushright}

\vspace{\baselineskip}

\begin{center}
\vspace{0.5\baselineskip}
\textbf{\Large
The bottom-quark mass from non-relativistic\\[0.1cm]
sum~rules at NNNLO
}
\\
\vspace{3\baselineskip}
{\sc M.~Beneke$^a$, A.~Maier$^a$, J.~Piclum$^{a,b,c}$, T.~Rauh$^a$}\\
\vspace{0.7cm}
{\sl ${}^a$Physik Department T31, James-Franck-Stra\ss{}e~1,\\
Technische Universit\"at M\"unchen, D-85748 Garching, Germany\\[0.3cm]
${}^b$Institut f\"ur Theoretische Teilchenphysik und Kosmologie,\\
  RWTH Aachen University, D-52056 Aachen, Germany\\[0.3cm]
${}^c$Albert Einstein Center for Fundamental Physics,\\
Institute for Theoretical Physics,
Sidlerstrasse 5,\\
CH-3012 Bern,
Switzerland}

\vspace*{1.2cm}
\textbf{Abstract}\\
\vspace{1\baselineskip}
\parbox{0.9\textwidth}{
We determine the mass of the bottom quark from high moments of the
$b\overline{b}$ production cross section in $e^+ e^-$ annihilation,
which are dominated by the threshold region. On the theory side
next-to-next-to-next-to-leading order (NNNLO) calculations both for the
resonances and the continuum cross section are used for the first
time. We find $m_b^\text{PS}(2\,\text{GeV}) =
4.532^{+0.013}_{-0.039}\,\text{GeV}$ for the potential-subtracted mass
and $m_b^{\MS}(m_b^{\MS}) = 4.193^{+0.022}_{-0.035}\,$GeV for the
$\overline{\rm MS}$ bottom-quark mass.
}
\end{center}

\newpage
\setcounter{page}{1}

\section{Introduction}
\label{sec:intro}

The bottom-quark mass is one of only a handful of fundamental QCD
parameters, and thus its precise knowledge is of considerable interest
by itself. Furthermore, there are also phenomenological applications
which benefit from a small uncertainty in the value of the bottom-quark
mass. Examples include Higgs decays to bottom quarks and decays of $B$
mesons.

Sum rules provide a well-established method for the determination of
heavy-quark masses~\cite{Novikov:1976tn,Novikov:1977dq}.
Considering the normalised bottom production cross section
\begin{equation}
  \label{eq:Rb_def}
  R_b(s) = \frac{\sigma(e^+e^- \to b\overline{b} + X)}{\sigma(e^+e^- \to
    \mu^+ \mu^-)}\,,
\end{equation}
where $s$ is the square of the $e^+e^-$ center-of-mass energy,
we can define its moments ${\cal M}_n$ as
\begin{equation}
\label{eq:Mn_def}
{\cal M}_n = \int_0^\infty ds\,\frac{R_b(s)}{s^{n+1}}
= - 6 \pi i \oint_{{\cal C}} ds\,\frac{\Pi_b(s)}{s^{n+1}}
= \frac{12 \pi^2}{n!} \bigg(\frac{d}{dq^2}\bigg)^n
\Pi_b\big(q^2\big)\bigg|_{q^2 = 0}\,.
\end{equation}
$\Pi_b$ denotes the contribution from bottom quarks to the vacuum
polarisation function. ${\cal C}$ is an arbitrary closed contour that
encloses the origin and does not cross the branch cut, i.e. the part of
the positive real half-axis where $R_b(s) = 12 \pi \im \Pi_b(s+i\epsilon) >0$.
Quark-hadron duality now permits to determine the bottom-quark mass by
equating moments obtained from the measured hadronic
cross section with moments calculated from derivatives of the
theoretically predicted polarisation function, which can themselves
be obtained from the theoretically predicted partonic cross section by
the above dispersion relation~(\ref{eq:Mn_def}).

For small values of $n$ it is most convenient to calculate moments
directly from the
derivatives of the polarisation function in conventional perturbation
theory. As $n$ increases the theory uncertainty grows and for $n \gtrsim
10$ the perturbation series shows no signs of convergence anymore
(cf.~\cite{Boughezal:2006uu,Maier:2007yn}).

For larger values of $n$ the integral over $R_b$ in
equation~(\ref{eq:Mn_def}) is increasingly dominated by small
kinetic energies $E =
\sqrt{s} - 2m_b \sim m_b/n$~\cite{Voloshin:1987rp,Voloshin:1995sf},
where $m_b$ is the bottom-quark mass. Thus, the quarks are
non-relativistic with a small velocity $v \sim 1/\sqrt{n} \ll 1$. At the
same time the strong coupling constant $\alpha_s$ is of the same order
as the quark velocity. For this reason terms scaling with powers of
$\alpha_s/v$, which originate from Coulomb interaction, have to be
summed to all orders. This is achieved in the effective theory of
potential non-relativistic QCD (PNRQCD)~\cite{Pineda:1997bj}. In summary,
for large-$n$, ``non-relativistic'' moments
the power counting for the cross section up to NNNLO is given by
\begin{equation}
  \label{eq:PNRQCD_power_counting}
  R_b \sim v \sum_k \bigg(\frac{\alpha_s}{v}\bigg)^{\!k}\times
  \begin{cases}
    1 & \text{LO}\\
    \alpha_s,v & \text{NLO}\\
    \alpha_s^2,\alpha_s v, v^2 & \text{NNLO}\\
    \alpha_s^3,\alpha_s^2 v,\alpha_s v^2, v^3 & \text{NNNLO}
  \end{cases}\,.
\end{equation}
The admissible values of $n$ are limited by the requirement that the
ultrasoft scale $m_b/n$ remains larger than the intrinsic strong
interaction scale of a few hundred MeV~\cite{Beneke:1999fe}.

In this work we present the first determination of the bottom-quark mass
from large-$n$ sum rules at next-to-next-to-next-to-leading order (NNNLO).
The corresponding NNLO calculations have been done about fifteen
years ago~\cite{Beneke:1999fe,Melnikov:1998ug,Hoang:1999ye}. A partial
NNNLO result that uses NNNLO accuracy for the bound-state contribution
and NNLO accuracy for the continuum contribution to the moments
appeared recently~\cite{Penin:2014zaa}. The outline of this paper
is as follows.  First, in
section~\ref{sec:moments_exp}, we derive the values of the relevant
moments from experimental data. We proceed in
section~\ref{sec:moments_th} with an outline of the corresponding PNRQCD
calculation and a discussion of suitable mass schemes. In
section~\ref{sec:num_analysis} we summarise our results for $R_b$ and
for the bottom-quark mass and compare them to other recent
high-precision sum rule analyses. For the comparison we consider the
works by Chetyrkin et al.~\cite{Chetyrkin:2009fv}, Hoang et
al.~\cite{Hoang:2012us}, and Penin and Zerf~\cite{Penin:2014zaa}. We
conclude in section~\ref{sec:conclusions}. Since we are aiming at
high precision, we include the effects of the non-zero charm-quark mass.
The details of this computation are given in four appendices.

\section{Experimental moments}
\label{sec:moments_exp}

For sufficiently large values of $n$ the experimental moments ${\cal
M}_n^\text{exp}$ are dominated by $\Upsilon$ bound states. In the
narrow-width approximation for the resonances $\Upsilon(1S)$ to
$\Upsilon(4S)$ we obtain
\begin{equation}
  \label{eq:Mn_exp_res}
  {\cal M}_n^\text{exp} = 9\pi \sum_{N=1}^4
  \frac{1}{\alpha(M_{\Upsilon(NS)})^2}\frac{\Gamma_{\Upsilon(NS)\to l^+
      l^- }}{M_{\Upsilon(NS)}^{2n+1}}
  + \int_{s_\text{cont}}^\infty ds\, \frac{R_b(s)}{s^{n+1}}\,,
\end{equation}
where we take the current PDG values~\cite{Beringer:1900zz} for the
masses $M_{\Upsilon(NS)}$ and the leptonic widths
$\Gamma_{\Upsilon(NS)\to l^+l^-}$ of the $\Upsilon$ resonances. In the
energy region of interest we approximate the running QED coupling by a
constant value, which is given in terms of the fine structure constant
by $\alpha(2m_b) \approx 1.036\, \alpha$~\cite{Jegerlehner:2011mw}.

For the remaining continuum integral we use experimental
data~\cite{Aubert:2008ab} corrected for initial state radiation between
$\sqrt{s_\text{cont}} = 10.6178\,\text{GeV}$ and $11.2062\,\text{GeV}$
(see~\cite{Chetyrkin:2009fv} for details). In the absence of data for
higher center-of-mass energies we assume $R_b$ to stay roughly constant
with $R_b=0.3 \pm 0.2$. For $n=6,10,15$ the unknown high-energy part
constitutes approximately $53\%, 35\%, 21\%$ of the total continuum
contribution, respectively. The large uncertainty of this part is
therefore not expected to be important.  The resulting values and
uncertainties for the experimental moments are shown in
table~\ref{tab:Mn_exp}.

\begin{table}
\centering
\begin{tabular}{l*{5}c}
\toprule
$n$        & 6          & 7          & 8          & 9          & 10  \\
\midrule
resonances & 0.1861(20) & 0.2004(22) & 0.2166(24) & 0.2351(27) & 0.2560(29) \\
continuum  & 0.0240(85) & 0.0182(58) & 0.0140(41) & 0.0110(29) & 0.0088(21) \\
total      & 0.2101(88) & 0.2185(62) & 0.2307(47) & 0.2461(39) & 0.2648(36) \\
\bottomrule
\end{tabular}

\vspace{1.5em}

\begin{tabular}{l*{5}c}
    \toprule
    $n$        & 11         & 12         & 13         & 14         & 15         \\\midrule
    resonances & 0.2797(33) & 0.3064(36) & 0.3364(40) & 0.3702(45) & 0.4081(50) \\
    continuum  & 0.0070(15) & 0.0057(11) & 0.0046(8)\phantom{0} & 0.0038(6)\phantom{0} & 0.0031(4)\phantom{0} \\
    total      & 0.2867(36) & 0.3120(38) & 0.3410(41) & 0.3740(45) & 0.4112(50) \\
\bottomrule
\end{tabular}
\caption{Contributions to the experimental moments ${\cal
      M}_n^\text{exp}$ in $(10\,\text{GeV})^{-2n}$.}
\label{tab:Mn_exp}
\end{table}

\section{Theory moments}
\label{sec:moments_th}

\subsection{The cross section in PNRQCD}
\label{sec:Rb_PNRQCD}

The normalised cross section for bottom production at NNNLO in
PNRQCD has the form
\begin{equation}
  \label{eq:Rb_theo}
  R_b = 12 \pi e_b^2 \im\bigg[\frac{N_c}{2m_b^2}\bigg(c_v\bigg[c_v -\frac{E}{m_b}\bigg(c_v+
  \frac{d_v}{3}\bigg)\bigg]G(E)+ \dots\bigg)\bigg]\,,
\end{equation}
see e.g.~\cite{Beneke:2013jia}.
$e_b$ and $m_b$ are the bottom quark's fractional electric charge and
pole mass, respectively. $E$ is the kinetic energy with the usual
relation $E = \sqrt{s} - 2m_b$ to the center-of-mass energy
$\sqrt{s}$. $c_v$ and $d_v$ are the matching coefficients of the
non-relativistic vector current, i.e.
\begin{equation}
  \label{eq:j_matching}
  j^i = c_v \psi^\dagger \sigma^i\chi + \frac{d_v}{6m_b^2}\psi^\dagger
  \sigma^i{\bf D}^2\chi+\dots\,,
\end{equation}
where $j^i$ are the spatial components of the relativistic current
$\bar b\gamma^\mu b$, and
$\psi$ ($\chi$) the non-relativistic quark (antiquark) spinors. Finally,
$G(E)$ is the non-relativistic current correlator
\begin{equation}
  \label{eq:Green_fun_def}
  G(E) = \frac{i}{2N_c(d-1)}\int d^dx\, e^{iEx^0} \langle 0|T [\chi^\dagger\sigma^i\psi](x)[\psi^\dagger\sigma^i\chi](0)|0\rangle\Big|_\text{PNRQCD}
\end{equation}
in $d$ space-time dimensions.

Below threshold, i.e. for $E<0$, $G(E)$ develops infinitely many poles
which can be interpreted as S-wave, spin-1 $b\bar b$ bound states.
In the vicinity of such a bound state $G(E)$ behaves as
\begin{equation}
  \label{eq:G_res}
  G(E) \xrightarrow{E \to E_N} \frac{|\psi_N(0)|^2}{E_N - E - i\epsilon}\,,
\end{equation}
where $\psi_N(0)$ is the bound state wave function at the origin while
$E_N$ is the binding energy.
Splitting off the bound-state contribution we can thus write the moments as
\begin{equation}
  \label{eq:Mn_th}
    {\cal M}_n^\text{th} = \frac{12\pi^2 N_c e_b^2}{m_b^2}
    \sum_{N=1}^\infty\frac{Z_N}{(2m_b + E_N )^{2n+1}}
  + \int_{4 m_b^2}^\infty ds\,\frac{R_b(s)}{s^{n+1}}
\end{equation}
with the residues
\begin{equation}
  \label{eq:Zn_def}
Z_N = c_v\bigg[c_v -\frac{E_N}{m_b}\bigg(c_v+
  \frac{d_v}{3}\bigg)\bigg] |\psi_N(0)|^2+ \dots\,.
\end{equation}
Note that $Z_N$ is a physical quantity, while the wave function $\psi_N(0)$
and the matching coefficients separately depend on the factorisation
scale and scheme that separates short- and long-distances.

$R_b$~(\ref{eq:Rb_theo}) and $Z_N$~(\ref{eq:Zn_def}) are understood to
be strictly expanded up to NNNLO according to the PNRQCD power
counting~(\ref{eq:PNRQCD_power_counting}). For example, terms of order
$\alpha_s^4 $ and higher that are generated through the product of
lower-order terms are to be dropped. Contrarily, we will leave the
factor $(2m_b + E_N)^{-(2n+1)}$ in equation~(\ref{eq:Mn_th}) unexpanded
for reasons discussed in section~\ref{sec:mass_schemes}. Not expanding
this factor is equivalent to resumming the poles back into the
polarisation function (cf.~\cite{Beneke:1999qg}), i.e. replacing
\begin{equation}
  \label{eq:pole_resum}
  \Pi_b \rightarrow \Pi_b + \frac{3e_b^2}{2m_b^2} \sum_{N=1}^\infty\bigg\{
  \frac{\big[Z_N\big]_{\text{expanded}}}
  {\big[E_N - E\big]_{\text{unexpanded}}} -
  \bigg[\frac{Z_N}{E_N - E}\bigg]_{\text{expanded}}\bigg\}
\end{equation}
in the contour integral that defines the moments~(\ref{eq:Mn_def}).

\subsection{Technical implementation}
\label{sec:technical}

We require the following ingredients for the NNNLO analysis of
non-relativistic moments:
\begin{itemize}
\item The hard Wilson coefficient $c_v$ at order
$\alpha_s^3$~\cite{Marquard:2014pea} and
$d_v$ at order $\alpha_s$~\cite{Luke:1997ys}.
\item The third-order corrections to the S-wave energy levels
$E_N$~\cite{Kniehl:2002br,Penin:2005eu,Beneke:2005hg}, to
the wave functions $\psi_N$ at the
origin~\cite{Penin:2005eu,Beneke:2005hg,Beneke:2007gj}
and to $G(E)$~\cite{Beneke:2005hg,Beneke:2014} from non-relativistic
potentials up to NNNLO~\cite{Beneke:2013jia,Wuester:2003,Kniehl:2001ju,Beneke:2014qea,Anzai:2009tm,Smirnov:2009fh}.
\item Ultrasoft corrections to these
quantities~\cite{Kniehl:1999ud,Beneke:2007pj,Beneke:2008cr}.
\end{itemize}
For an extensive overview of all third-order corrections
see~\cite{Beneke:2013jia,Beneke:2014}.

Note that the results
in~\cite{Beneke:2008cr,Beneke:2014} are calculated for the case of top
quarks, that is in the complex energy plane for finite imaginary
part $\Gamma$ corresponding to the top-quark width.
The application of these results to bottom quarks
requires to take the limit $\Gamma\to 0$, which is non-trivial,
since the analytic expressions cannot be straightforwardly evaluated
numerically for vanishing width. In~\cite{Beneke:2008cr}
the $\Gamma = 0$ result has been constructed for the ultrasoft
contribution by extrapolation. Evaluating the third-order potential
corrections given in~\cite{Beneke:2014} for real energy $E$ requires
a substantial amount of extra work, which we briefly discuss in the
following.

The higher-order potential corrections to the Green function are expressed in
terms of nested harmonic sums, sums over gamma and polygamma functions,
and generalised hypergeometric functions~\cite{Beneke:2005hg,Beneke:2014}.
The complex variable $\lambda
= (\alpha_s C_F/2) \sqrt{-m/(E + i\Gamma)}$ appears for example in the
argument of (poly-)gamma functions or as one of the parameters of the
hypergeometric functions. In our application, we have to evaluate the
Green function for positive values of the energy $E$, starting at
$E=0$. For vanishing width $\Gamma$, $\lambda$ tends to $+i\infty$ as
$E$ tends to zero. Thus, we have to ensure that all expressions are
well-defined in this limit and that their numerical evaluation is
possible.

In most cases it is possible to express the correction to the Green
function in terms of harmonic sums. This is in particular the case for
most of the generalised hypergeometric functions, which can be treated
as described in appendix~A.1 of~\cite{Beneke:2011mq}. The harmonic sums
can then be analytically continued with the methods
of~\cite{Blumlein:2009ta,Albino:2009ci}. However, in some cases the
correction to the Green function is expressed in terms of single or even
double sums which could not be expressed as nested harmonic sums. For
such sums it was often necessary to truncate the summation at some
($\lambda$-dependent) value and construct suitable asymptotic expansions
to approximate the remainder. In all cases we have checked that the
numerical precision is sufficient for our extraction of the bottom-quark
mass, such that the numerical uncertainty can be neglected.

A relatively simple example of this procedure is given by the sum
\begin{equation}
\sum\limits_{k=1}^\infty \frac{\big[(k - \lambda)
\big(\psi(k - \lambda) -\psi(k)\big) + k \lambda
\psi^{(1)}(k)\big]^2}{k} \,,
\label{eq:example}
\end{equation}
which appears in the insertion of the Darwin term. Here
$\psi$ is the logarithmic derivative of the gamma function and
$\psi^{(1)}$ the first derivative of $\psi$. The sum converges only
slowly when $\lambda$ is large, which makes the numerical evaluation
difficult. Therefore, we introduce a cut-off $\Lambda$ for the
summation and explicitly sum all terms up to this cut-off. Choosing
$\Lambda$ to be much larger than $|\lambda|$, we can
approximate the remainder by expanding the summand in the limit
$k\to\infty$. Note that for $\psi(k - \lambda)$ this is not simply an
expansion in $|\lambda|/k$, but rather a double expansion for $k \gg
1$ and $k \gg |\lambda|$. In the first step the entire argument of the
$\psi$ function is considered large and the terms of the resulting
asymptotic series are further expanded for $|\lambda|/k \to 0$ in the
second step, yielding
\begin{align}
\psi(k-\lambda) ={}& \ln(k-\lambda) -
\frac{1}{2(k-\lambda)} - \frac{1}{12(k-\lambda)^2} +
\frac{1}{120(k-\lambda)^4} + {\mathcal
O}\left(\frac{1}{(k-\lambda)^6}\right) \nonumber \\
 ={}& \ln k - (1 +
2\lambda) \frac{1}{2k} - (1 + 6\lambda + 6\lambda^2) \frac{1}{12k^2} -
(\lambda + 3\lambda^2 + 2\lambda^3) \frac{1}{6k^3} \nonumber \\
 & + (1 - 30\lambda^2 - 60\lambda^3 - 30\lambda^4) \frac{1}{120k^4} +
{\mathcal O}\left(\frac{1}{k^5}\right) \,.
\end{align}
After expanding the summand in (\ref{eq:example}),
the sum over $k$ from $\Lambda+1$ to
infinity can be evaluated in terms of Hurwitz zeta functions (in more
complicated cases we also encounter derivatives of this function). The
first three terms are
\begin{equation}
 \frac{\lambda^4}{4}\, \zeta(3, \Lambda + 1) +
\frac{\lambda^5}{6}\, \zeta(4, \Lambda + 1) + \frac{\lambda^4}{36}
\left(-3 + 4\lambda^2\right) \zeta(5, \Lambda + 1)\,,
\end{equation}
where $\zeta(s,a) = \sum_{k=0}^\infty (k + a)^{-s}$. Due
to the nature of the double expansion of $\psi(k -\lambda)$, higher
order terms in this series can have the same powers of $\lambda$ as
lower order ones. However, these terms are still suppressed by
additional powers of $\Lambda$, since the first argument of $\zeta$
increases with each term and $\zeta(s,\Lambda)$ with $s>1$ behaves like
$1/\Lambda^{s-1}$ as $\Lambda$ tends to infinity. In practice, we fix
for each sum an appropriate expansion depth $N_{\rm max}$ for the
remainder. This makes it possible to speed-up the calculation by
pre-computing the expansion for arbitrary values of the cut-off. The
latter is then determined at runtime as
$\Lambda=\max(\Lambda_0,f|\lambda|)$, where $\Lambda_0$ is a lower limit
for the cut-off and $f$ is a positive integer. $\Lambda_0$ and $f$ are
again chosen separately for each sum. In the above example we use
$N_{\rm max}=15$, $\Lambda_0=100$, and $f=2$. By varying $N_{\rm max}$
and $f$, we can check the numerical stability of the procedure.

\subsection{Mass schemes}
\label{sec:mass_schemes}

It is well known that for quark masses the pole scheme is ambiguous due
to its sensitivity to infrared
renormalons~\cite{Beneke:1994sw,Bigi:1994em,Beneke:1994rs}.
Numerous short-distance mass schemes have been developed to cure this
shortcoming~\cite{Bigi:1996si,Beneke:1998rk,Hoang:1998ng,Pineda:2001zq}. In
this work we consider the potential-subtracted (PS) mass introduced
in~\cite{Beneke:1998rk} and the mass in the \MS{} scheme. They are
related to the pole mass via perturbative series of the form
\begin{equation}
  \label{eq:rel_pole_mass}
m_b = m_b^M + \sum_{i=0}^\infty \delta
m^M_i\,,\qquad M = \text{PS},\MS\,.
\end{equation}
For the cancellation of leading infrared renormalons we have to
take into account the first correction term $\delta m^M_0$ already at
leading order and one additional correction term for each further order
in the PNRQCD expansion.

In the PS scheme the leading subtraction term is given by
\begin{equation}
  \label{eq:delta_m_0_PS} \delta m_0^\text{PS} =
\frac{\alpha_s(\mu)C_F }{\pi}\,\mu_f\,,
\end{equation}
 where we choose the subtraction scale as $\mu_f =
2\,\text{GeV}$. The choice of the renormalisation scale $\mu$ is
discussed in section~\ref{sec:mu_choice}. The higher-order terms up to
$\delta m^\text{PS}_3$ required for our NNNLO analysis can be found
in~\cite{Beneke:2005hg}.

In the \MS{} scheme the higher-order corrections in
relation~(\ref{eq:rel_pole_mass}) are only known up to
$i=2$~\cite{Chetyrkin:1999qi,Melnikov:2000qh}.\footnote{Note that in
our notation $\delta m_i^{\MS}$ denotes the $(i+1)$-loop correction.} It
is, however, expected that the correction
terms $\delta m^{\MS}_i$ are dominated by the leading infrared
renormalon already at relatively low orders~\cite{Beneke:1994qe,Ball:1995ni}.
On this basis an approximation was constructed in~\cite{Pineda:2001zq,Ayala:2014yxa}.
The deviation from the known result is about $10\%$ for $i=1$
and less than $1\%$ for $i=2$. We employ this approximation for $i=3$,
which corresponds to setting the correction term $\delta m^{\MS}_3$ at
the scale $m^{\MS}_b$ to $\tilde{r}_3m_b^{\MS} \alpha_s^4\approx
13.59(83)\, m_b^{\MS} \alpha_s^4$ in the notation
of~\cite{Pineda:2001zq}. The error range encompasses the
value $\tilde{r}_3= 13.5972$ obtained from the large-$\beta_0$
approximation~\cite{Beneke:1994qe,Ball:1995ni}.

When eliminating the pole mass in the cross section formulae in favour
of a short-distance mass the question arises whether the resulting
expression should be expanded in the correction terms $\delta m_i^M,\, i
\geq 1$. For the \MS{} scheme such an expansion is
not sensible in the threshold region~\cite{Beneke:1998rk}. This can for
instance be seen by considering the factor $(2 m_b+E_N)^{-(2n+1)}$ in
the resonance contribution to the theory moments~(\ref{eq:Mn_th}). On
the one hand, the correction of order $i$ to the energy levels $E_N$ is
parametrically of the same order as $\delta m_{i+1}^{\MS}$ according to
the PNRQCD power counting $\alpha_s \sim v \ll 1$. On the other hand,
renormalon cancellation only occurs between the correction of order $i$
to the binding energies and $\delta m_i^{\MS}$, so an expansion
consistent with our power counting would spoil this cancellation.

In the PS scheme both expanding and not expanding the PS-pole
mass relation~(\ref{eq:rel_pole_mass}) appear to be viable
options. However, as discussed in~\cite{Beneke:2014}, an expansion
induces unphysical behaviour near threshold in the continuum cross
section. Therefore we will not expand the cross section in
$\delta m_i^\text{PS}$, which corresponds to the PS-shift (PSS)
prescription of~\cite{Beneke:2014}. At the same
time this again implies that the entire factor $(2 m_b+E_N)^{-(2n+1)}$ must
remain unexpanded to ensure the cancellation of leading infrared
renormalons. In practice, this means that in both schemes, for given
$m_b^M$ and order N$^k$LO, we first compute $m_b$ from
(\ref{eq:rel_pole_mass})  truncating the sum at
$i=k$ and add the bound state energy evaluated to the same
order.

\subsection{Expansion in the kinetic energy}
\label{sec:Rb_res}

Up to now, in~(\ref{eq:Rb_theo}), (\ref{eq:Mn_th}) and
(\ref{eq:Zn_def}) an overall factor of $1/s$ stemming from the
relativistic polarisation function has been expanded for $E\ll m_b$.
Since we chose not to expand the factor $1/s^{n+1}$ in the
definition~(\ref{eq:Mn_def}) of the moments, we may also contemplate to
keep this factor unexpanded. The corresponding expressions, denoted by a
tilde, take the following forms:
\begin{align}
  \label{eq:Rb_unexp}
  \widetilde{R}_b &= 12 \pi e_b^2 \im\bigg[\frac{2 N_c}{s}\bigg(c_v\bigg[c_v -\frac{E}{m_b}
  \frac{d_v}{3}\bigg]G(E)+ \dots\bigg)\bigg]\,,\\
  \label{eq:Z_unexp}
\widetilde{Z}_N &= c_v\bigg[c_v - \frac{E_N}{m_b} \frac{d_v}{3}\bigg]
|\psi_N(0)|^2+ \dots\,,\\
  \label{eq:Mn_unexp}
    {\cal \widetilde{M}}_n^\text{th} &= 48\pi^2 N_c e_b^2
    \sum_{N=1}^\infty\frac{\widetilde{Z}_N}{(2m_b + E_N)^{2n+3}}
  + \int_{4 m_b^2}^\infty ds\,\frac{\widetilde{R}_b(s)}{s^{n+1}}\,.
\end{align}
Formally, the difference to~(\ref{eq:Rb_theo}),
(\ref{eq:Mn_th}) and (\ref{eq:Zn_def}) is of higher order (NNNNLO).

These higher-order corrections are numerically non-negligible, possibly
due to sub-leading renormalon contributions. For the
$N$th resonance $1/s$ assumes the form $(2m_b + E_N)^{-2}$, where
renormalon contributions cancel between the binding energy and the
mass. If we now expand this factor for $E_N\ll m_b$ and, according to
the PNRQCD power counting, discard contributions to $E_N$ that are
beyond NLO the renormalon cancellation will again be spoilt. Since
the exponent now is not of order $n$ as was the case for the $1/s^{n+1}$
factor, the generated renormalon ambiguity is only of order
$\Lambda_{\rm QCD}/m_b$ and therefore sub-leading. Since this is not
the only source of sub-leading renormalon contributions,
in order to decide which prescription for the expansion
of $1/s$ is the preferred one it would be necessary to analyse carefully
how all of these contributions can be cancelled in the determination of
the moments. To our knowledge such an analysis has not been performed
yet. We will find in
section~\ref{sec:mb_det} that not expanding the factor $1/s$ in the
polarisation function improves the consistency of mass values extracted
from different moments. In the following, we will therefore mainly
concentrate on the ``unexpanded'' moments defined by~\eqref{eq:Mn_unexp}.
We then check that the difference to the
``expanded'' approach is compatible with our estimate of the
perturbative uncertainty.

\subsection{Charm-quark mass effects}
\label{sec:charm_mass}

The ingredients for the NNNLO cross section described in
section~\ref{sec:Rb_PNRQCD} all assume the light quarks to be massless,
which is well justified as long as the light-quark masses are smaller
than the physical scales that appear in the problem.  However the actual
charm-quark mass is of the order of the soft scale (inverse Bohr
radius) $m_b\alpha_s\sim m_b
v$ of the $b\bar{b}$ system near threshold and therefore has to be
included in a consistent treatment. At NNLO the effects of a non-zero
charm-quark mass on the moments have been discussed thoroughly
in~\cite{Hoang:2000fm}, where they were found to lead to a sizeable
shift of around $-30\,$MeV in the extracted \MS{} bottom-quark mass.

Since the factor $1/s^{n+1}$ in the moments is expanded
non-relativistically in~\cite{Hoang:2000fm} the results cannot be
included in our expressions~\eqref{eq:Mn_th} and \eqref{eq:Mn_unexp}.
In the following we discuss in which steps in the computation of the
cross section the effects of a non-zero charm-quark mass are relevant
and determine the missing contributions up to NNLO. The NNNLO charm-mass
effects are unknown at the time of this writing and their determination
is clearly outside the scope of this work.

The computation of the cross section proceeds in three separate steps,
which are the hard matching, the soft matching and the computation of
the spectral function in the resulting effective theory PNRQCD, as was
discussed in detail in~\cite{Beneke:2013jia}.  Since $m_c$ is considered
to be soft the results of the hard matching must be analytic in
$m_c^2/m_b^2\sim\alpha_s^2$ and charm-mass effects due to the insertion
of a charm loop into a gluon line scale as
$\alpha_s^2m_c^2/m_b^2\sim\alpha_s^4$ compared to $\alpha_s$ for the
gluon line without charm-loop insertion.
To NNNLO the hard matching procedure is therefore unaffected.

In the soft matching procedure the charm quark is integrated out. The only
part of the resulting PNRQCD Lagrangian that is affected at NNLO is the
Coulomb potential, which can be split in two parts
\begin{equation}
 V=V_\text{massless}+V_{m_c},
\end{equation}
where $V_{m_c}$ is defined such that it vanishes for
$m_c\rightarrow0$.\footnote{ The effects of integrating out the charm
quark on the running of $\alpha_s$ can be included in the Coulomb
potential. We follow the convention of~\cite{Hoang:2000fm}, where the
potential is defined for $\alpha_s$ evolving with
$n_{l,\text{massless}}+1$ flavors.}  It has been computed to two loops
in~\cite{Melles:1998dj,Melles:2000dq}; convenient representations are
given in~\cite{Hoang:2000fm}. In order to offer a self-contained
discussion we quote these results in
appendix~\ref{sec:charm_potential}. As discussed in~\cite{Beneke:2013jia}
the soft matching of the external vector current with massless light
quarks is trivial to all orders, because the respective integrals are
scaleless. The introduction of the charm-quark mass as a soft scale
means that starting at two loops this is no longer true.  However the
corresponding matching coefficient must be trivial in the limit
$m_c\rightarrow0$ and since the only other scale relevant for the
external current is $m_b$, the correction must contain at least one
power of $m_c/m_b$, which again is beyond NNLO.\footnote{ We have also
checked explicitly that the term without any power of $m_c/m_b$ cancels
upon wave function renormalisation.}

The spectral function receives contributions from the charm-quark
mass through insertions of the Coulomb potential. At
NLO only the single insertion of $V_{m_c}^{(1)}$ is required. At NNLO
the single insertion of $V_{m_c}^{(2)}$ and the double insertions of
$V_{m_c}^{(1)}$ and $V_\text{massless}^{(1)}$ as well as twice
$V_{m_c}^{(1)}$ contribute. Our results for the charm corrections to the
energy levels $E_N$ and the wave functions $|\psi_N(0)|^2$ can be found
in appendix~\ref{sec:charm_psi}. We find numerical
agreement in those parts that are available in the
literature~\cite{Eiras:2000rh, Hoang:2000fm}.

Since, for the reasons discussed in section~\ref{sec:mass_schemes},
we use the PS or $\overline{\text{MS}}$ mass instead of the pole mass
in the cross section as well as the moments, the charm-mass effects
also have to be considered in the relation
between different mass schemes. In the conversion between the pole and
the \MS{} scheme, these effects are known at order
$\alpha_s^3$~\cite{Bekavac:2007tk}. For our analysis we have also computed
the corresponding corrections to the relation between the PS and the
pole mass. These results are summarised in appendix~\ref{sec:charm_PS/pole}.

Anticipating our numerical analysis in section~\ref{sec:num_analysis}
let us now discuss the impact of a non-zero charm-quark mass. We
parametrise the corrections in terms of the \MS{} charm-quark mass at
our overall renormalisation scale $\mu$, which we obtain via 4-loop
running from the initial value
$m_c^{\MS}(3\,\text{GeV})=0.986\,$GeV~\cite{Chetyrkin:2009fv}.\footnote{To
compute the renormalisation group evolution we employ either
\texttt{RunDec}~\cite{Chetyrkin:2000yt} or a custom implementation.} It
should be noted that effects from a non-zero charm-quark mass are
expected to be large for quantities that have large infrared
sensitivity. The reason for this is that the charm-loop correction
effectively acts as an infrared cut-off on the virtuality of the gluon
line into which it is inserted.

\begin{table}
  \centering
  \begin{tabular}{lcccccc}
    \toprule
    $N=1$                     & $2\delta m_{1,m_c}^{\text{PS}}$ & $2\delta m_{2,m_c}^{\text{PS}}$ & $E_{1,m_c}^{(1)}$   & $E_{1,m_c}^{(2)}$    & $f_{1,m_c}^{(1)}$   & $f_{1,m_c}^{(2)}$  \\\midrule
    single insertion
    $\{m_c\}$                 & 9.50                     & 23.21                    & -6.82         & -18.02          & 0.0335           & 0.0644           \\
    double insertions         \\
    $\{m_c,\text{massless}\}$ & --                          & --                          & --               & -1.46           & --               & -0.0043           \\
    $\{m_c,m_c\}$             & --                          & --                          & --               & -0.02           & --               & -0.0001           \\[0.5cm]
    $\sum$                   & 9.50                      & 23.21                    & -6.82         & -19.50          & 0.0335           & 0.0600           \\\bottomrule
  \end{tabular}
  \caption{Contributions of a non-zero charm-quark mass to the binding
energy $E_{1,m_c}^{(i)}=E_{1}^{(0)}e_{1,m_c}^{(i)}$ and wave function of
the $\Upsilon(1S)$ resonance. The corrections to the wave function are
given by $|\psi_1(0)|^2 = |\psi_1^{(0)}(0)|^2\big(1+\sum_{i=1}^\infty
f_1^{(i)}\big)$, and $f_{1,m_c}^{(i)}$ denotes the charm correction to
$f_1^{(i)}$. The results are given in the pole-mass scheme with the
input values $m_b=5\text{ GeV}$, $\mu=5\text{ GeV}$, $\alpha_s(5\text{
GeV})=0.2135$ and $m_c(5\text{ GeV})=0.892\text{ GeV}$. To demonstrate
the large cancellation in the $\Upsilon(1S)$ mass
$M_{\Upsilon(1S)}=2 m_b+E_1= 2m_b^{\text{PS}} + 2\sum_{i=1}^\infty\delta m_i^{\text{PS}} +
E_1$ once a short-distance mass scheme is used,
we also show results for the charm-mass effects in the relation between
the PS and pole mass with $\mu_f=2\text{ GeV}$.
$2\delta m_{i,m_c}^{\text{PS}}$ and $E_{1,m_c}^{(i)}$ are given in
MeV, while $f_{1,m_c}^{(i)}$ is dimensionless.}
  \label{tab:charm_effects}
\end{table}

Table~\ref{tab:charm_effects} illustrates the numerical impact of a
non-zero charm-quark mass on the first energy level and the
corresponding wave function at the origin and demonstrates the
cancellations in the transition to a short-distance mass scheme. Let us
now discuss the effect on the final value of the \MS{} bottom-quark mass
extracted from the tenth moment. If, at first, we only consider the
corrections from single insertions to the binding energies and the wave
functions at NNLO (NLO) the resulting mass value receives a shift of
$+16\,$MeV ($+4\,$MeV). In line with our previous discussion we expect a
considerable compensation from the charm effects in the relation between
the infrared-sensitive pole mass and the PS mass. In fact, including
also these corrections the mass shift reduces to around $+0.5\,$MeV
($-0.5\,$MeV). The cancellation of infrared contributions can also be
observed analytically. Expanding all corrections in the limit of a small
charm-quark mass, the linear terms in the expansion cancel exactly
between the corrections to the energy levels and wave functions on the
one side and the relation between PS and pole mass on the other side
(cf.~\cite{Hoang:2000fm}). The charm mass corrections to the relation
between the PS and the \MS{} scheme are relatively small; adding them to
the analysis leads to a total shift of $-3.5\,$MeV ($-2\,$MeV) in the
\MS{} bottom-quark mass.

Up to now, we have neglected charm corrections to double insertions and
to the continuum cross section. In agreement with~\cite{Hoang:2000fm} we
find that the charm contributions from double insertions are suppressed
compared to the single insertions, causing an additional mass shift of
only $+0.5\,$MeV. Since the overall continuum contribution to the tenth
moment is already small we expect charm effects in the cross section
above threshold to be negligible. Indeed we find an extra shift of less
than $0.1\,$MeV from the NLO corrections listed in
appendix~\ref{sec:charm_greenfunctionl}. In the light of these findings
we will use the computationally expensive charm corrections to the
continuum cross section and the double insertions only to extract our
central value and neglect them in the error analysis.

To account for the unknown NNNLO charm-mass corrections we assign an
uncertainty given by the difference between the bottom-quark mass
obtained at NNLO for the physical value of the charm-quark mass and the
bottom-quark mass obtained in the limit of a vanishing charm-quark mass.

Concerning the size of the charm corrections there is a large
discrepancy between the final mass shift of $-3\,$MeV in our analysis
and about $-30\,$MeV in~\cite{Hoang:2000fm}. We find that
differences between the two approaches, such as the choice of
renormalisation scales, expansion of factors $1/s$, and different values
for the charm-quark mass, cannot account for this disparity.
It seems suspicious
that~\cite{Hoang:2000fm} claims large effects of around $50\%$ for
$n=20$. Such high moments are dominated by the first resonance. By the
definition of the $1S$ mass, however, charm effects cancel completely in
the combination $2m_b^{1S}+2\sum_{i=1}^\infty \delta m_i^{1S}+E_1 \equiv 2m_b^{1S}$. Thus,
only the charm corrections to the residue $Z_1$ contribute. These
effects are independent of $n$, and, according to our findings, rather
moderate in size (cf. table~\ref{tab:charm_effects}).

\section{Numerical analysis}
\label{sec:num_analysis}

\subsection{Choice of the renormalisation scale}
\label{sec:mu_choice}

\begin{figure}
  \centering
  \includegraphics[width=.72\textwidth]{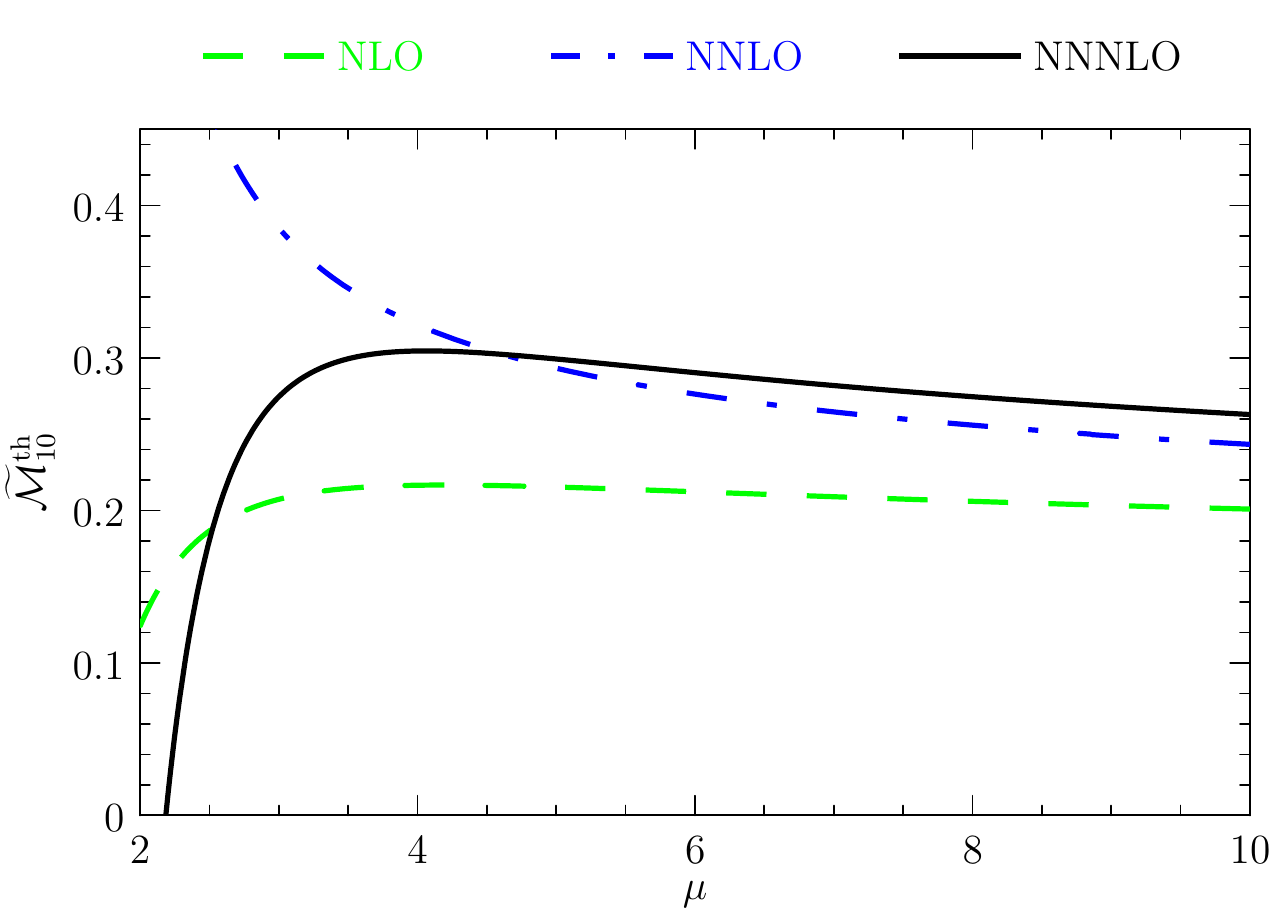}
  \caption{Scale dependence of the tenth moment $\widetilde{\cal
M}_{10}^\text{th}$ in $(10\,\text{GeV})^{-20}$ for $\alpha_s(M_Z) = 0.1184$ and
$m_b^\text{PS}(2\,\text{GeV}) = 4.5\,$GeV.}
  \label{fig:M10_scale}
\end{figure}

Since the moments receive contributions from several distinct physical
regions it is important to choose an adequate value for the overall
renormalisation scale $\mu$. A priori ``natural'' options include the
hard scale $\mu \sim m_b$, the soft scale $\mu \sim 2m_b/\sqrt{n}$ and
the ultrasoft scale $\mu \sim m_b/n$. In this work we do not consider the
possibility of summing logarithmic effects $\ln \sqrt{n}$ related
to the presence of different scales by renormalisation group methods
(see~\cite{Hoang:2012us} in the present context), since
$\ln\sqrt{10}$ is not a particularly
large number. Figure~\ref{fig:M10_scale} shows
the scale dependence of the tenth moment for $\mu$ between $2\,$GeV and
$10\,$GeV at different orders. There is clearly no convergence of the
perturbation series for $\mu \lesssim 3\,$GeV.\footnote{A similar
observation was made for the leptonic decay width of the $\Upsilon(1S)$
resonance~\cite{Beneke:2014qea}.} We therefore adopt $\mu =
m_b^{\text{PS}}$ as our central scale and vary $\mu$ between $3\,$GeV
and $10\,$GeV to estimate the uncertainty. Note that there is no overlap
in the moment values at NLO and NNLO over the entire scale range,
which is one of the motivations to perform the third-order calculation.
The figure shows that the NNNLO curve lies within the interval determined
by the NNLO scale variation.

\subsection{Comparison to the fixed-order continuum}
\label{sec:comp_cont}

The present work is the first to include the continuum cross section
with NNNLO accuracy (heavily relying on the input
from~\cite{Beneke:2008cr,Beneke:2014} as described in
section~\ref{sec:technical}), which is also the most complicated
part of the NNNLO calculation. Since sum rule determinations of
the bottom-quark mass rely on quark-hadron duality, the continuum
is conceptually a crucial ingredient in the analysis.

\begin{figure}[t]
  \centering
  \includegraphics[width=.72\textwidth]{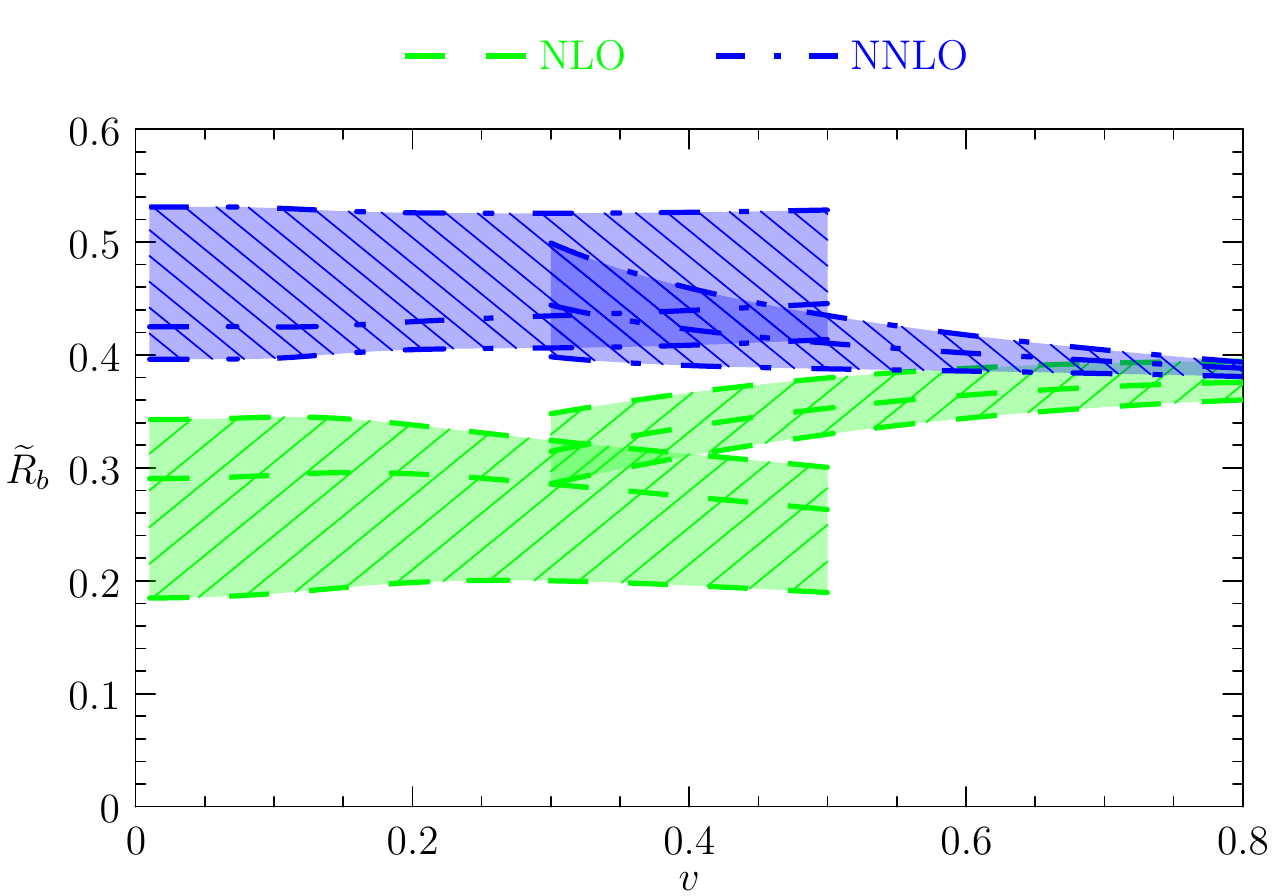}\\
  \includegraphics[width=.72\textwidth]{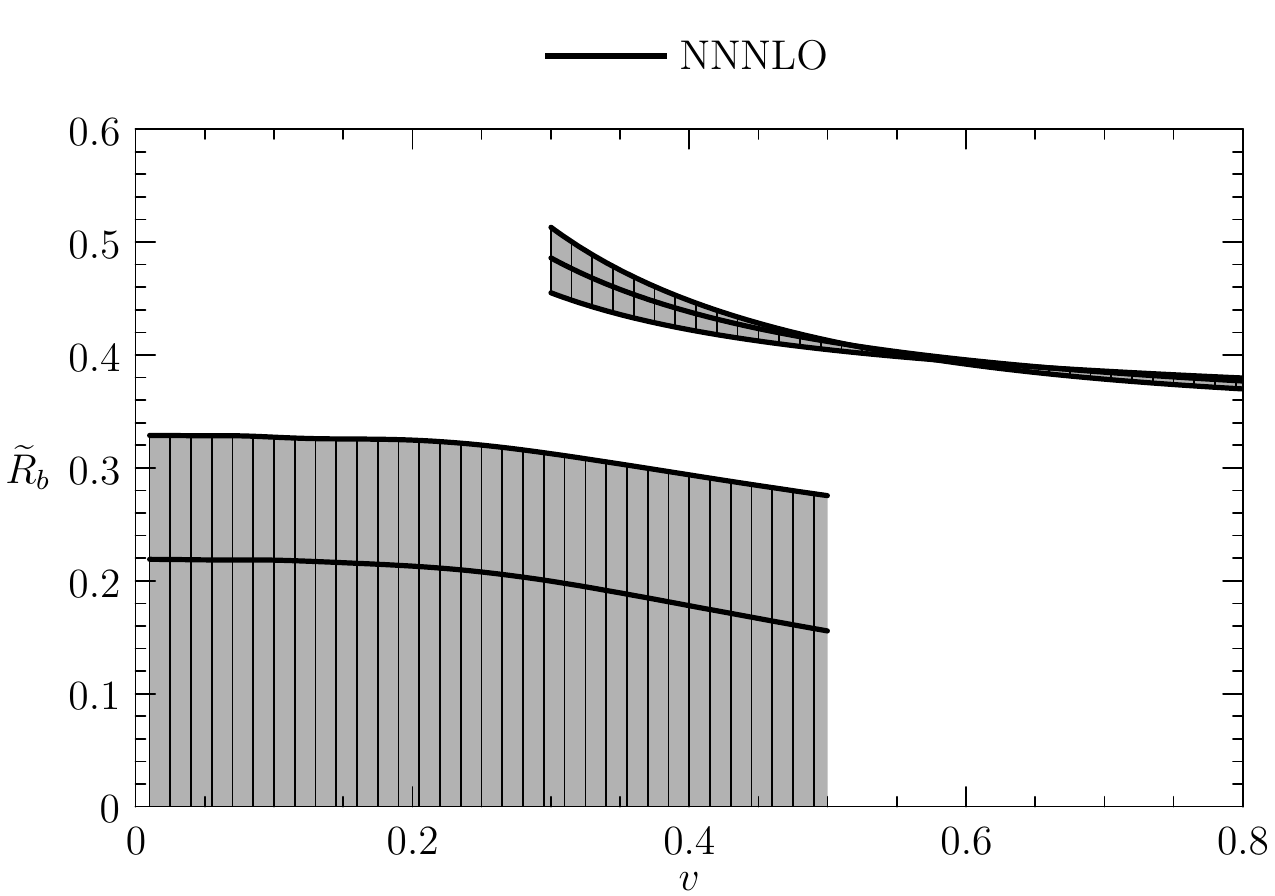}
  \caption{Behaviour of $\widetilde{R}_b$ as a function of $v =
\sqrt{E/m_b}$ for a pole mass of $m_b = 5\,$GeV. The curves on the left
show the PNRQCD prediction, whereas the curves on the right correspond
to fixed-order perturbation theory. The shaded areas show the
uncertainty from varying the renormalisation scale between $3\,$GeV and
$10\,$GeV.}
  \label{fig:cont}
\end{figure}

The summation of factors $\alpha_s/v$ implicit in the PNRQCD calculation
of the correlator $G(E)$ is only necessary close to
threshold. It is therefore reasonable to expect that there is a region with
$\alpha_s \ll v \ll 1$ where the continuum cross section can be calculated
reliably both in PNRQCD (requiring $v\ll 1$) and conventional
perturbation theory (requiring $v\gg\alpha_s$). In
figure~\ref{fig:cont} we show the respective predictions adopting the
pole-mass scheme with $m_b = 5\,$GeV, and without expanding the factor
$1/s$ in the polarisation function (see section~\ref{sec:Rb_res}). For
the fixed-order curves we have used the analytically known result up to
order $\alpha_s$~\cite{Kallen:1955fb} and Pad\'e
approximation~\cite{Chetyrkin:1996cf,Hoang:2008qy,Kiyo:2009gb} at orders
$\alpha_s^2$ and $\alpha_s^3$.

At NLO and NNLO (upper panel) there is apparently a good agreement
between the two theories for $v \sim 0.4$. The NNNLO curve in
PNRQCD (lower panel), however, lies
significantly below the NNLO and the fixed-order curves and shows a very
strong dependence on the renormalisation scale.\footnote{Note that for
the $\Upsilon(1S)$ resonance the behaviour of the NNNLO correction under scale
variation is much better, see~\cite{Beneke:2014qea}.} For smaller
scales $\mu$ we even observe unphysical negative values for
$\widetilde{R}_b$.

One reason for the considerable difference between fixed-order QCD and
PNRQCD at NNNLO may be the limited information from the threshold region
used for the Pad\'e approximation. In particular, the behaviour of the
NNNLO fixed-order curve for $v \to 0$ is by construction determined by
the na\"ive expansion of the NNLO PNRQCD polarisation function to order
$\alpha_s^3$~\cite{Hoang:2008qy,Kiyo:2009gb}. Although they are
suppressed by an additional factor of $v$, corrections from NNNLO PNRQCD
could alter this behaviour considerably as the NNNLO PNRQCD result
differs significantly from the NNLO one.

One example for such a big missing correction is given by the product of
the third-order correction to the hard Wilson coefficient and the
leading-order $G(E)$. Parametrically this correction is of order
$v\,\alpha_s^3$. Due to the numerically large factors involved it still
causes a shift of $\Delta R_b \sim -0.2$ at $v = 0.4$. At higher orders
in $\alpha_s$ there will be further sizeable contributions from this
product, e.g. an additional negative shift of $-0.2$ at order
$\alpha_s^4v^0$.

While it may be possible to reconcile the discrepancy between the
relativistic fixed-order and PNRQCD predictions there still remains the
problem that the convergence of the continuum cross section in
resummed non-relativistic perturbation theory
appears to be quite poor. Neither the difference between consecutive
orders nor the scale uncertainty shrink when considering higher-order
corrections to the continuum cross section. In fact, at NNNLO the scale
uncertainty is much larger than at any lower order.

\begin{figure}
  \centering
  \includegraphics[width=.72\textwidth]{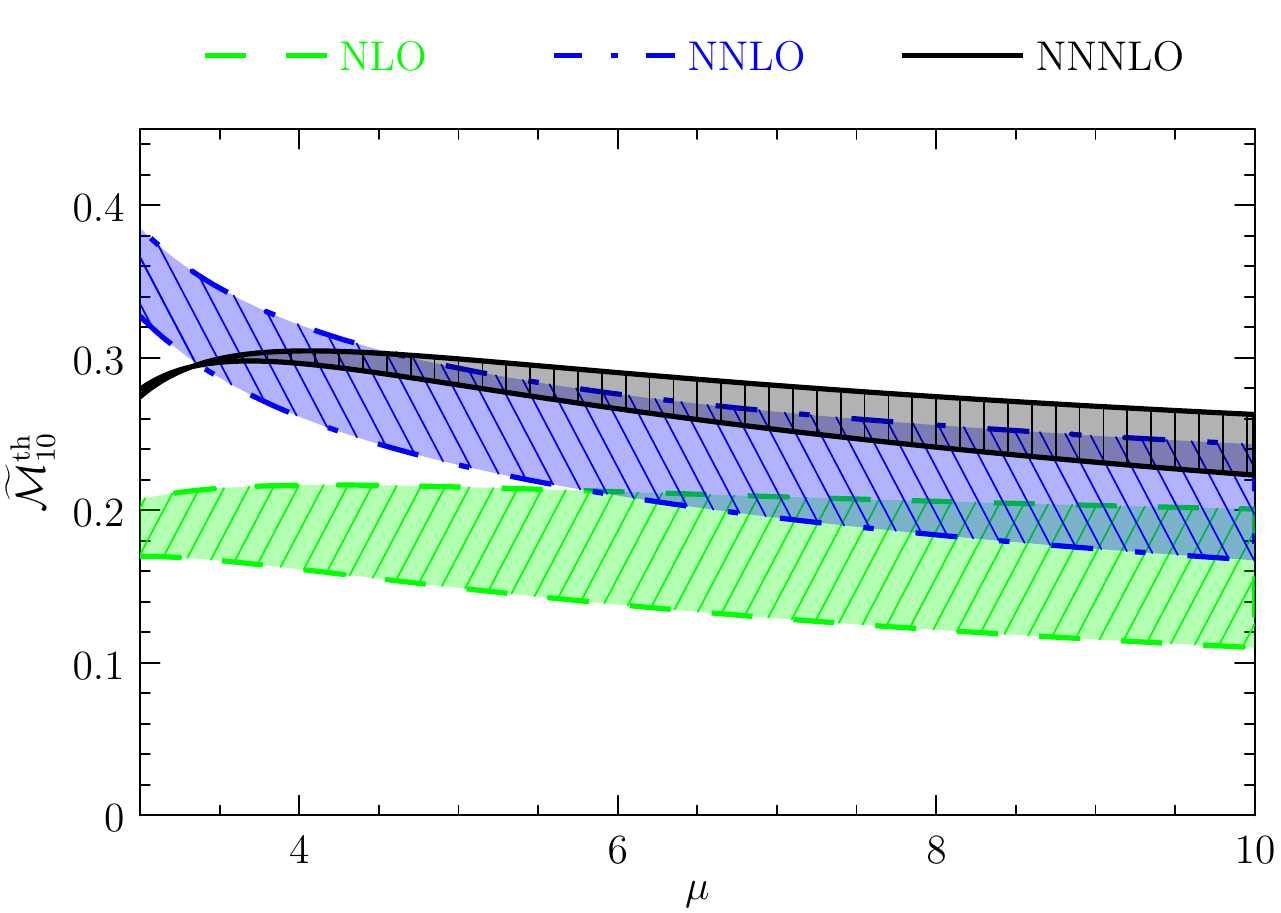}
  \caption{Scale dependence of $\widetilde{\cal M}_{10}^\text{th}$ in
$(10\,\text{GeV})^{-20}$ with $m_b^\text{PS}(2\,\text{GeV}) = 4.5\,$GeV.
At each
order the shaded band shows the contribution from the
continuum cross section. The lower boundary of the band is the moment
without the continuum contribution.}
  \label{fig:M10tilde_cont}
\end{figure}

In our analysis, however, we are of course not interested in the
continuum cross section itself but in non-relativistic moments. These
are expected to receive their dominant contribution from the
resonances. Figure~\ref{fig:M10tilde_cont} shows the continuum
contribution to the tenth moment at different orders in perturbation
theory as a function of the renormalisation scale $3\,\text{GeV} < \mu <
10\,\text{GeV}$. As expected we observe that the continuum contribution
at NNNLO is rather small, amounting to less than $5\%$ at our central
scale $\mu = m_b$ and about $15\%$ at $\mu = 10\,$GeV. Furthermore, the
continuum contribution reduces both the distance between
consecutive orders and the scale dependence at each order. We conclude
that the seemingly problematic behaviour of the continuum cross section
does not impede the extraction of the bottom-quark mass from large-$n$
moments at NNNLO.

\subsection{Determination of the bottom-quark mass}
\label{sec:mb_det}

We are now in a position to determine the bottom-quark mass by requiring
${\cal M}_n^\text{th} = {\cal M}_n^\text{exp}$ for moments with
$n\approx 10$. We first eliminate the pole mass in favour of the PS mass
$m_b^\text{PS}(\mu_f)$, and then convert the resulting mass to the \MS{}
scheme. Irrespective of the order of the moments we always perform the
conversion at order $\alpha_s^4$. As a part of our error analysis we
will also first convert to the \MS{} mass $m_b^{\MS}(\overline{\mu})$ at an
intermediate scale $\overline{\mu}$, which we keep separate from the
overall renormalisation scale $\mu$, and then use renormalisation group
evolution to find the scale-invariant mass $m_b^{\MS}(m_b^{\MS})$.

To estimate the error of the resulting mass value for a given moment
${\cal M}_n$ we consider the following sources of uncertainties.
\begin{itemize}
 \item Experimental uncertainty. We add in quadrature the
errors induced by uncertainties in the $\Upsilon$ masses, the leptonic
widths, the BaBar data directly above the
resonances~\cite{Aubert:2008ab}, and our estimate $0.1 \leq R_b \leq
0.5$ in the high-energy region.
\item Unknown higher-order corrections. As detailed in
section~\ref{sec:mu_choice} we choose the extracted PS mass as our
central renormalisation scale and vary $\mu$ between $3\,$GeV and
$10\,$GeV.
\item Uncertainty in the strong coupling. We evolve $\alpha_s(m_Z) =
0.1184$ down to our central scale using four-loop running, and decouple
the bottom quark at $\mu_{\text{thr}} = 2\*m_b$. The exact choice of the
decoupling scale is numerically irrelevant. When varying the
renormalisation scale we remain in the four-flavour theory. To determine
the uncertainty in the strong coupling we vary its value at the scale of
the $Z$ boson mass by $\pm 0.0010$.
\item QED effects. In our analysis, we include the leading correction in
the conversion to the $\MS$ mass scheme and the QED Coulomb
potential. Assuming $\alpha \sim \alpha_s^2$, these effects are formally
of NLO. In practice, we find a shift of less than $1\,$MeV in the
extracted quark mass.
\item Number of theoretical resonances. In practice we only take into
account the contribution from the first six resonances in the
formulae~(\ref{eq:Mn_th}), (\ref{eq:Mn_unexp}) for the theoretical
moments and estimate the resulting uncertainty from the difference
to considering only four resonances.
\item Scheme conversions. When extracting the PS mass, we vary $\mu_f$
between $1\,$GeV and $3\,$GeV. In the conversion to the \MS{}
scheme, we vary the intermediate scale $\overline{\mu}$ between $3\,$GeV
and $10\,$GeV and estimate the error from
the unknown value of the conversion coefficient $\delta m_3^{\MS}$ as
described in section~\ref{sec:mass_schemes}. All scheme conversion
errors are added in quadrature.
\item Charm-mass effects. As detailed in section~\ref{sec:charm_mass} we
  take the difference between the mass values at NNLO with and without
  charm effects.
\item Non-perturbative effects. To estimate the order of magnitude, we
follow~\cite{Voloshin:1995sf} (see also~\cite{Onishchenko:2000yy}) and
consider the leading contribution to the operator product expansion
(OPE) from the gluon condensate. It is assumed that
higher-dimensional condensates and higher-order corrections to the
Wilson coefficients can be neglected. Note that the OPE is only valid
for $m_b v^2\sim m_b/n \gg \Lambda_\text{QCD}$. The need to avoid an
uncontrolled, non-perturbative ultrasoft contribution limits the
admissible values of $n$. Under these assumptions
we find a negligible effect of less than $1\,$MeV on the value of the bottom
mass.

The smallness of the gluon condensate contribution to moments
of order $n=10$ is surprising, since the corresponding contribution
to the $\Upsilon(1S)$ is large (though uncertain, see the recent discussion
in~\cite{Beneke:2014qea}) and the high moments are
already completely determined by the $\Upsilon(1S)$. While a certain
amount of cancellation of non-perturbative effects is expected in the
moments due to quark-hadron duality, the degree of cancellation (about
one part in 500 for $n=10$) is somewhat puzzling. We therefore advocate
that the estimate of non-perturbative effects from the gluon condensate
correction is considered with some caution and refrain from using it
to limit the allowed values of $n$. More details on the gluon
condensate contribution are given in appendix~\ref{app:nonperturbative}.
\end{itemize}

\begin{figure}
  \centering
  \includegraphics[width=.72\textwidth]{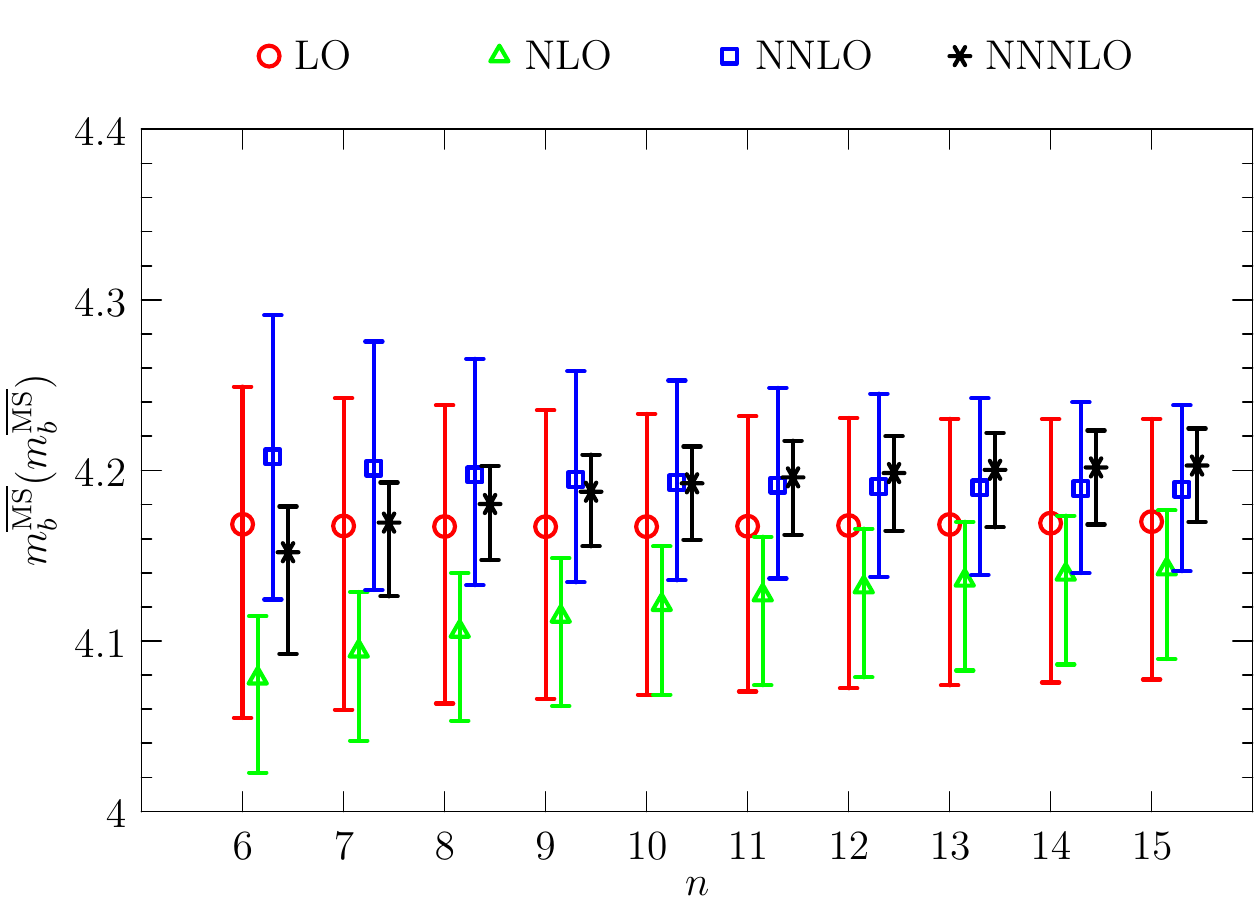}
  \caption{Values of $m_b^{\MS}(m_b^{\MS})$ in GeV obtained from $m_b^\text{PS}$
for different moments ${\cal M}_n$.}
\label{fig:mb_PSshift}
\end{figure}

\begin{figure}
  \centering
  \includegraphics[width=.72\textwidth]{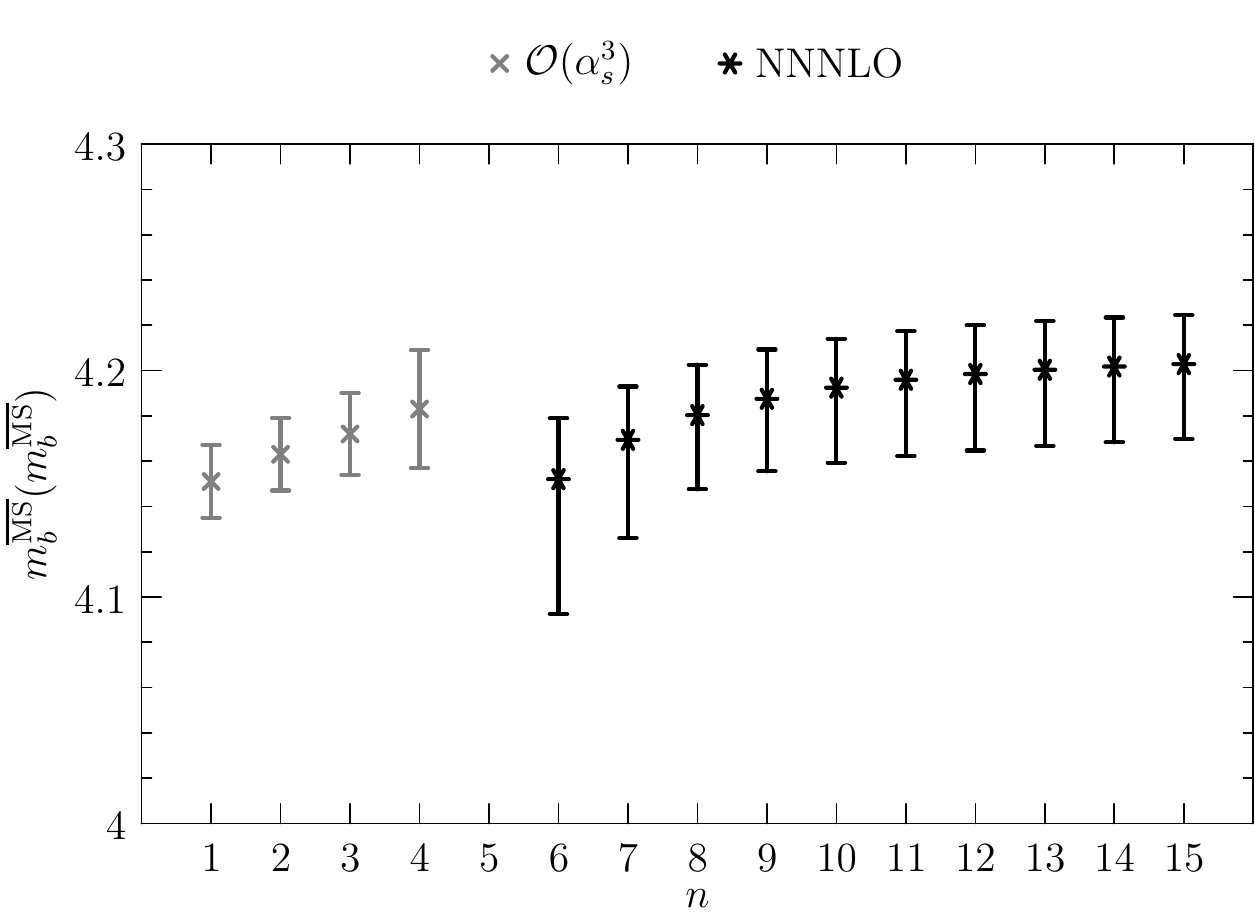} \\
  \includegraphics[width=.72\textwidth]{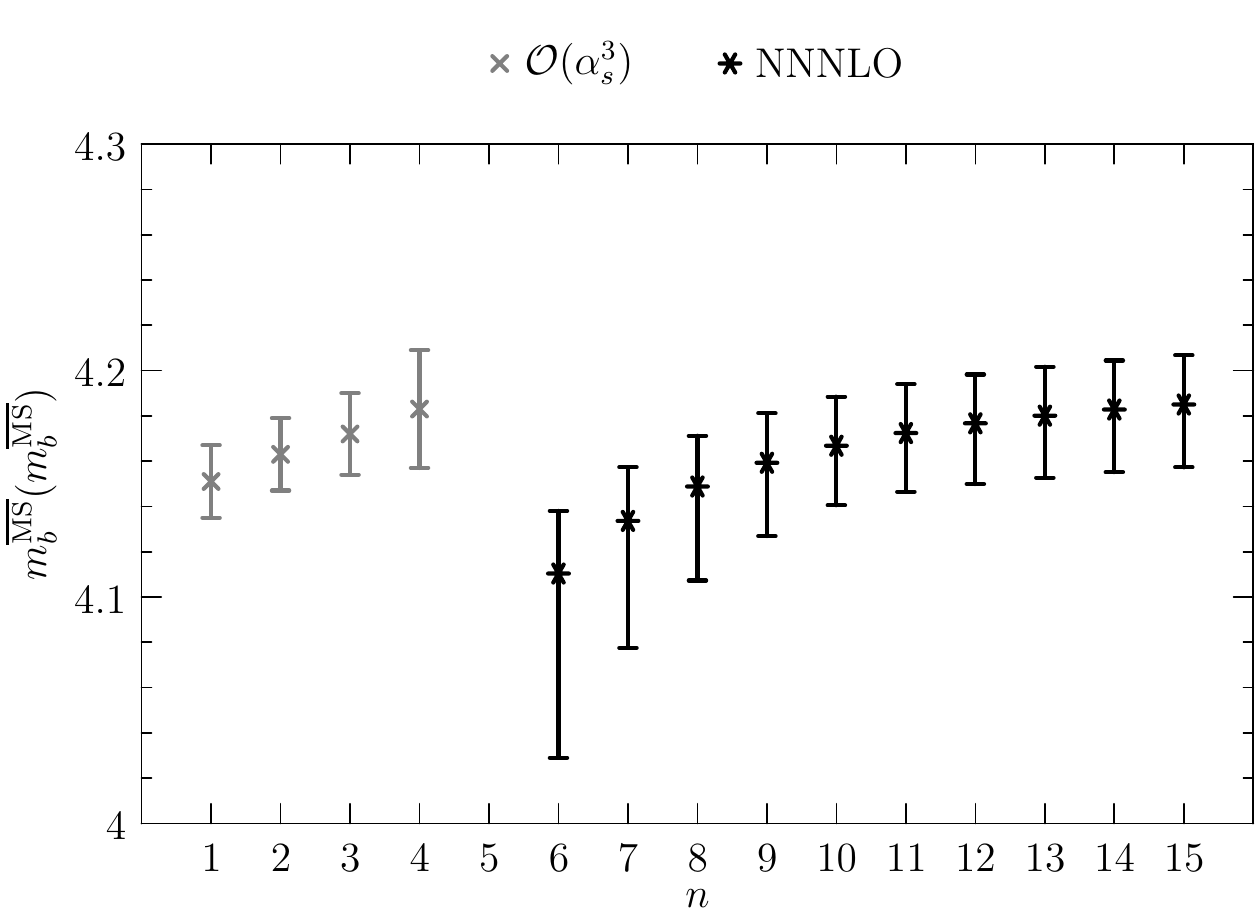}
  \caption{Values of $m_b^{\MS}(m_b^{\MS})$ in GeV obtained from
$m_b^\text{PS}$ for different NNNLO moments ${\cal M}_n$ with $ 6 \leq n
\leq 15$. In the upper panel we show the resulting mass values for
``unexpanded'' theoretical moments defined according to
(\ref{eq:Mn_unexp}); in the lower panel we have used
``expanded'' moments~(\ref{eq:Mn_th}).  To facilitate the
comparison to existing results we include the mass values obtained from
a fixed-order determination~\cite{Chetyrkin:2009fv} for $n\leq 4$. }
\label{fig:mb_PS_NNNLO}
\end{figure}

Our results for the \MS{} masses obtained from moments ${\cal M}_n$ with
$ 6 \leq n \leq 15$ are shown in figure~\ref{fig:mb_PSshift}. There
appears to be a good agreement of the mass values extracted from moments
with $n \approx 10$ and a reasonable convergence of the perturbation
series. For smaller values of $n$ the behaviour is considerably worse as
the non-relativistic approximation becomes less reliable. As can be seen
from figure~\ref{fig:mb_PS_NNNLO}, the mass values at NNNLO decrease due
to the small continuum cross section near threshold
(cf. section~\ref{sec:comp_cont}). The behaviour is worse
if ``expanded'' moments ${\cal M}_n$ are considered (lower panel).

As anticipated in section~\ref{sec:Rb_res} there is a considerable
difference between the mass values extracted from ``unexpanded'' moments
$\widetilde{\cal M}_n^\text{th}$ and ``expanded'' moments ${\cal
M}_n^\text{th}$. For the tenth moment the mass difference amounts to
$-26\,$MeV. Though quite large, this value still lies within our error
estimate for higher-order contributions obtained by varying the
renormalisation scale.

As our final result we adopt the PS and \MS{} masses extracted from
the 10th moment $\widetilde{\cal M}_{10}^\text{th}$ leading to
\begin{align}
  \label{eq:mass_res} m_b^\text{PS}(2\,\text{GeV}) ={}&
  \big[4.532^{+0.002}_{-0.035}(\mu) \pm 0.010 (\alpha_s)
   ^{+ 0.003}_{-0} (\text{res}) \pm 0.001 (\text{conv}) \notag\\
&\pm 0.002 (\text{charm}) ^{+
  0.007}_{-0.013} (n) \pm 0.003 (\text{exp})\big]\,\text{GeV}\notag\\
={}&4.532^{+0.013}_{-0.039}\,\text{GeV}
\,,\\
  \label{eq:mass_res_MS}
m_b^{\MS}(m_b^{\MS}) ={}& \big[4.193 ^{+ 0.002}_{-0.031} (\mu) \pm 0.001
(\alpha_s) ^{+ 0.003}_{-0} (\text{res}) ^{+ 0.021}_{-0.010} (\text{conv})\notag\\&
\pm0.002(\text{charm}) ^{+
  0.006}_{-0.012} (n) \pm 0.003 (\text{exp})\big]\,\text{GeV}\notag\\
={}&4.193^{+0.022}_{-0.035}\,\text{GeV}
\,.
\end{align}
In addition to the uncertainties discussed above we added the
differences to the central mass values obtained from $\widetilde{\cal
M}_{8}^\text{th}$ and $\widetilde{\cal M}_{12}^\text{th}$ to our error
estimate. The corresponding term is marked by the label $(n)$. Our value
for the \MS{} mass is in good agreement with determinations from
relativistic (small $n$) sum rules at order
$\alpha_s^3$~\cite{Chetyrkin:2009fv} and approximate NNNLO
non-relativistic sum-rules~\cite{Penin:2014zaa}. A more detailed
comparison is given in section~\ref{sec:compare}.

An alternative method is to forego the PS scheme and directly use
relation~\eqref{eq:rel_pole_mass} to eliminate the pole mass in favour
of the $\MS$ mass. For this ``direct'' extraction of the $\MS$ mass we
obtain
\begin{align}
  \label{eq:mass_res_i}
  m_b^{\MS}(m_b^{\MS}) ={}& \big[4.194 ^{+ 0.001}_{-0.026} (\mu) \pm 0.001
(\alpha_s) ^{+ 0.003}_{-0} (\text{res}) ^{+0.008}_{-0.010} (\text{conv})\notag\\&
\pm0.002(\text{charm}) ^{+
  0.007}_{-0.013} (n) \pm 0.003 (\text{exp})\big]\,\text{GeV}\notag\\
={}&4.194^{+0.012}_{-0.030}\,\text{GeV}\,.
\end{align}

\subsection{Comparison with previous works}
\label{sec:compare}

As can be seen from table~\ref{tab:comp_mb}, recent sum rule
determinations of the bottom-quark mass are in reasonably good
agreement.\footnote{We do not include NNNLO determinations of the
bottom-quark mass from the $\Upsilon(1S)$ mass, since the theoretical
uncertainty is dominated by non-perturbative effects (see
e.g.~\cite{Beneke:1999fe}) and is not competitive with those given in
table~\ref{tab:comp_mb}.} In the following we summarise the differences
between the listed analyses.
\begin{table}
  \centering
  \begin{tabular}{cccc}
\toprule
Analysis & Central moment & Perturbative order &
$m_b^{\MS}(m_b^{\MS})\,[\text{GeV}]$ \\\midrule
CKMMMSS~\cite{Chetyrkin:2009fv} & ${\cal M}_{2}$ & $\alpha_s^3$ &$4.163 \pm 0.016$ \\
 HRS~\cite{Hoang:2012us} & ${\cal M}_{10}$ & NNLO + NNLL &$4.235 \pm 0.055$ \\
 PZ~\cite{Penin:2014zaa} & ${\cal M}_{15}$ & approx. NNNLO & $4.169 \pm 0.009$ \\
 This work & ${\cal M}_{10}$ & NNNLO & $4.193^{+0.022}_{-0.035}$\\\bottomrule
  \end{tabular}
  \caption{Bottom-quark masses obtained from different sum rule
analyses. In the last column experimental and theoretical errors were
added in quadrature.}
\label{tab:comp_mb}
\end{table}

\subsubsection{Chetyrkin et al. (CKMMMSS)~\cite{Chetyrkin:2009fv}}
\label{sec:compare_CKMMMSS}

CKMMMSS consider relativistic moments ${\cal M}_n,\, n \le 4$ in
fixed-order perturbation theory to order $\alpha_s^3$. Our experimental
input largely corresponds to the one used by CKMMMSS. However, as low
moments are
much more sensitive to the experimental high-energy continuum, CKMMMSS
rely on the prediction from perturbation theory above $11.24\,$GeV and
use a linear interpolation between this region and the last data point
at $11.2062\,\text{GeV}$. CKMMMSS choose the same scale for the strong
coupling and the \MS{} mass and estimate the perturbative uncertainty
from correlated scale variation and an estimated upper bound for the
$\alpha_s^4$ coefficient. Since we vary the two scales independently,
the perturbative error estimates are not directly comparable. We expect
however that an independent scale variation generally yields a more
conservative error estimate. Conversely, if we set $\overline{\mu}
\equiv \mu$ in our analysis we obtain $m_b^{\MS}(m_b^{\MS}) =
4.194^{+0.012}_{-0.029}\,$GeV in place of (\ref{eq:mass_res_MS})
with a significantly smaller estimated uncertainty. The more
conservative error estimate using uncorrelated coupling and mass
renormalisation scale variations that should be compared to our result
(\ref{eq:mass_res_MS}) or (\ref{eq:mass_res_i}) is unfortunately not
available.

\subsubsection{Hoang, Ruiz-Femen\'ia and Stahlhofen
  (HRS)~\cite{Hoang:2012us}}
\label{sec:compare_HRS}
HRS perform a renormalisation group improved analysis of the moments ${\cal
M}_n$ with $6\le n \le 14$ and a default value of $n=10$ in the framework of
vNRQCD~\cite{Luke:1999kz}. In addition to the usual terms scaling as
$\alpha_s\sqrt{n}$ also logarithms $\alpha_s\ln \sqrt{n}$ are summed,
achieving NNLO + (partial) NNLL accuracy. Like CKMMMSS, HRS use the
perturbative QCD prediction in place of experimental data for the
high-energy continuum, but assign an uncertainty of $10\%$ to it. We
agree with the experimental moments used by HRS within the errors. HRS
first determine the so-called $1S$ mass, which is then converted to the
\MS{} scheme. The analysis of HRS does not include charm-mass effects,
which are currently unknown in their renormalisation group improved
framework.

HRS assume an uncertainty of $15\,$MeV in the conversion to the $\MS$
mass. This is comparable to our estimate, albeit slightly lower. Like
HRS we find that the dependence on $\alpha_s$ is reduced by the
conversion from the intermediate mass scheme to the $\MS$ scheme and is
quite small in the final result.

The central values of our analysis at NNLO and NNNLO almost coincide,
but lie about $45\text{ MeV}$ below the NNLL and about $80\text{ MeV}$
below the NNLO result obtained in HRS.
The analysis of HRS differs from ours in a number of aspects. HRS use the
1S instead of the PS mass, expand the factor $1/s^{n+1}$ non-relativistically
and expand the bound-state poles, the 1S-pole mass relation and the
factor $1/s$ in the vacuum polarisation function. They choose the
moment-dependent central scale $\mu=m_b(1/\sqrt{n}+0.2)$ instead of
$\mu=m_b$, but use the hard scale for the vector current matching
coefficients. Furthermore, HRS neglect charm-mass effects and use the
\MS-pole mass relation at order $\alpha_s^3$. We find that it is mainly the
scale choice which is responsible for the difference between the NNLO
results given by HRS and ours. The remaining differences in
the analyses have only moderate numerical effects, though their precise
size depends on the order in which they are implemented. As a net result,
when we adapt our analysis to the one of HRS, we reproduce their NNLO value
up to a negligible difference of $2\text{ MeV}$.
We note that the estimate of the perturbative uncertainty
at NNLO based on the scale variation of HRS is about twice as large as
the one used by us. Given that the convergence of the successive
approximations from NNLO to NNNLO at the scale $\mu=m_b$ is very good
(see figure~\ref{fig:M10_scale}), while low scales generally seem to
lead to a break-down of non-relativistic perturbation
theory~\cite{Beneke:2005hg}, we conclude that our error estimate is
sufficiently conservative.

The final result of HRS refers to the (partial)
NNLL calculation. The resummation of logarithms reduces the scale
dependence compared to the NNLO result and therefore compensates part of the
large positive shift of the bottom-quark mass introduced by choosing a
lower scale, leading to the bottom-quark mass quoted in
table~\ref{tab:comp_mb}. A previous less complete NNLO + partial NNLL
analysis~\cite{Pineda:2006gx} obtained $m_b^{\MS}(m_b^{\MS}) = 4.19 \pm 0.06$.

\subsubsection{Penin and Zerf (PZ)~\cite{Penin:2014zaa}}
\label{sec:compare_PZ}

PZ also use non-relativistic
sum rules, but choose even higher moments with a central value of
$n=15$. Such moments are rather insensitive to the
experimental high-energy continuum, which in their work is
neglected, but potentially introduce an unspecified systematic
uncertainty from ultrasoft effects that may already be in the
non-perturbative regime (further discussion in
appendix~\ref{app:nonperturbative}). Choosing $n=15$
instead of $n=10$ increases the resulting
mass value by approximately $12\,$MeV.

On the theory side, PZ employ the complete NNNLO PNRQCD prediction for
the resonances, and an estimate of $R_b = \rho\,
Z_1^{\text{NNNLO}}/Z_1^{\text{NNLO}} R^\text{NNLO},\,0.5\le\rho\le 2,$
for the continuum to account for the (then) unknown NNNLO contribution.
Our calculation shows that the true result lies somewhat below the band
spanned by the variation of the auxiliary parameter $\rho$.  Using
rescaled NNLO instead of NNNLO for the continuum leads to an increase of
about $5\,$MeV in the mass extracted from ${\cal M}_{15}$.  Not
expanding the factor $1/s$ in the polarisation function
(cf. section~\ref{sec:Rb_res}), however, again lowers the resulting mass
value by about $17\,$MeV. PZ directly extract the $\MS$ mass without any
intermediate threshold mass. As can be seen by comparing
(\ref{eq:mass_res_i}) and (\ref{eq:mass_res_MS}) the effect on the final
value is relatively small. It should be noted that PZ estimate a large
charm effect based on~\cite{Hoang:2000fm}, amounting to a mass shift of
about $-25\,$MeV, while our calculation results in a shift of only about
$-3\,$MeV (see discussion in section~\ref{sec:charm_mass}). Neglecting
QED and charm effects in both analyses the central values for
$m_b^{\MS}(m_b^{\MS})$ coincide ($4.195\,$GeV vs. $4.194\,$GeV).

Notwithstanding the similarities between the two analyses, PZ claim a
much smaller overall uncertainty of only 9 MeV in the value of the bottom-quark
mass. The main reasons for this are overly optimistic
estimates of the perturbative uncertainty and the uncertainty assigned
to the conversion between pole and $\MS$ mass, where PZ assume $2.1$ and
$2.2$\,MeV, respectively, resulting in a significant overestimate of
the final precision of the bottom-quark mass.

\section{Conclusions}
\label{sec:conclusions}

We have presented the first complete NNNLO determination of the
bottom-quark mass from non-relativistic sum rules.  We find a mass of
$m_b^\text{PS}(2\,\text{GeV}) = 4.532^{+0.013}_{-0.039}$ in the PS
scheme, which corresponds to an \MS{} mass of $m_b^{\MS}(m_b^{\MS}) =
4.193^{+0.022}_{-0.035}$.  Compared to previous NNLO analyses
of non-relativistic moments we observe
a significantly reduced uncertainty. This reduction is mostly due to the
choice of a higher central scale $\mu = m_b^{\text{PS}}$, which follows
from better insight into the convergence of successive approximations
which are now known up to NNNLO. In spite of
poor behaviour of the continuum cross section we observe that the NNNLO
moments are stable under scale variation, and moments with different
$n\approx 10$ are in good agreement with each other. Our results agree
reasonably well with other recent determinations of the bottom-quark mass
from the inclusive $e^+ e^-\to b\bar b$ cross section,
including NNNLO fixed-order analyses based on relativistic sum-rules.
Conservative uncertainty estimates of the bottom-quark \MS{} mass have now
reached the $\pm\,$(25--30)$\,$MeV range.

\subsection*{Acknowledgements}
\label{sec:acknowledgements}

This work has been supported by the DFG
Sonder\-forschungs\-bereich/Transregio~9 ``Com\-pu\-ter\-gest\"utzte
Theoretische Teil\-chen\-physik'', the Gottfried Wilhelm Leibniz
programme of the Deutsche Forschungsgemeinschaft (DFG), and the
Munich Institute for Astro- and Particle
Physics (MIAPP) of the DFG cluster of excellence "Origin and Structure of
the Universe". Some analytical calculations were performed with the help of
FORM~\cite{Vermaseren:2000nd}.

\appendix

\section{Charm effects in the Coulomb potential}
\label{sec:charm_potential}

To determine the effects of a non-zero charm-quark mass we split
the Coulomb potential into two parts
\begin{equation}
  \label{eq:Coulomb_split}
  \tilde{V} = \tilde{V}_\text{massless} + \tilde{V}_{m_c}\,,
\end{equation}
where $\tilde{V}_\text{massless}$ refers to the Coulomb
potential for a massless charm quark, so that the charm correction
$\tilde{V}_{m_c}$ vanishes for $m_c = 0$.

The corrections to the Coulomb potential due to the charm-quark mass
have been determined at NNLO in~\cite{Melles:1998dj,Melles:2000dq}.
We use the following dispersion relation representations
from~\cite{Hoang:2000fm}, which are very convenient for computations.
In momentum space the potential is given by
\begin{align}
\tilde{V}_{m_c}(\mathbf{q}) &= \sum_{i=1}^\infty\delta\tilde{V}_{m_c}^{(i)}(\mathbf{q})\,,\\
 \delta\tilde{V}_{m_c}^{(1)}(\mathbf{q})&=
-\frac{4\pi\alpha_sC_F}{\mathbf{q}^2}\frac{\alpha_s}{3\pi}T_F\left[\Pi(\mathbf{q}^2)-\left(\ln\frac{\mathbf{q}^2}{m_c^2}-\frac53\right)\right]\,,
 \label{eq:deltaV1tilde}\\
 \delta\tilde{V}_{m_c}^{(2)}(\mathbf{q})&=-\frac{4\pi\alpha_sC_F}{\mathbf{q}^2}\left(\frac{\alpha_s}{4\pi}\right)^2
 \begin{aligned}[t]
   \Bigg\{&\frac{8T_F}{3}\left[\Pi(\mathbf{q}^2)-\left(\ln\frac{\mathbf{q}^2}{m_c^2}-\frac53\right)\right]\left(a_1-\beta_0\ln\frac{\mathbf{q}^2}{\mu^2}\right)\\
   &+\left(\frac{4T_F}{3}\right)^2\left[\Pi(\mathbf{q}^2)-\left(\ln\frac{\mathbf{q}^2}{m_c^2}-\frac53\right)\right]^2\\
   &+\frac{76T_F}{3}\left[\Xi(\mathbf{q}^2)-\left(\ln\frac{\mathbf{q}^2}{m_c^2}-\frac{161}{114}-\frac{26}{19}\zeta_3\right)\right]\Bigg\}\,,
 \end{aligned}
 \label{eq:deltaV2tilde}
\end{align}
where
\begin{align}
  a_1 &= \frac{31}{9} C_A - \frac{20}{9}T_Fn_l\,,\\
  \beta_0 &= \frac{11}{3} C_A - \frac{4}{3}T_Fn_l\,,\\
 \Pi(\mathbf{q}^2)&=2\mathbf{q}^2\int\limits_1^\infty dx\frac{f(x)}{\mathbf{q}^2+4x^2m_c^2},\\
 \Xi(\mathbf{q}^2)&=2c_1\mathbf{q}^2\int\limits_{c_2}^\infty \frac{dx}{x}\frac{1}{\mathbf{q}^2+4x^2m_c^2}
                  +2d_1\mathbf{q}^2\int\limits_{d_2}^\infty \frac{dx}{x}\frac{1}{\mathbf{q}^2+4x^2m_c^2},\\
 f(x)&=\frac{\sqrt{x^2-1}}{x^2}\left(1+\frac{1}{2x^2}\right).
\end{align}
The parameters $c_1,c_2,d_1,d_2$ are adopted from~\cite{Hoang:2000fm} as
\begin{align}
  \label{eq:cd_1}
  c_1 &= \frac{\ln\frac{A}{d2}}{ \ln\frac{c2}{d2}}\,, \qquad\quad
  d_1 = \frac{\ln\frac{c2}{A}}{ \ln\frac{c2}{d2}}\,,\\
\label{eq:cd_2}
  c_2 &= 0.470\,, \qquad\quad \hspace*{0.14cm}d_2 = 1.120\,,\\
\label{eq:A}
  A &= \exp\bigg(\frac{161}{228} + \frac{13}{19}\zeta_3 -\ln
  2\bigg)\,.
\end{align}
The potential in configuration space can be obtained by a
Fourier transformation. We obtain
\begin{align}
 \delta V_{m_c}^{(1)}(r)={}&-\frac{C_F\alpha_s}{r}\frac{\alpha_s}{3\pi}\left[\int\limits_1^\infty dxf(x)e^{-2m_crx}+\left(\ln(\tilde{m}_cr)+\frac56\right)\right],
  \label{eq:deltaV1}\displaybreak[0]\\
 \delta V_{m_c}^{(2)}(r) ={}& -\frac{C_F\alpha_s}{r}\left(\frac{\alpha_s}{3\pi}\right)^2\Bigg\{\Bigg[-\frac32\int\limits_1^\infty dxf(x)e^{-2xm_cr}\nonumber\\
                       & \times\left(\beta_0\left[\ln\frac{4x^2m_c^2}{\mu^2}-\text{Ei}(2xm_cr)-e^{4xm_cr}\text{Ei}(-2xm_cr)\right]-a_1\right)\nonumber\\
                       & +3\left(\ln\left(\tilde{m}_cr\right)+\frac56\right)\left(\beta_0\ln\left(\tilde{\mu} r\right)+\frac{a_1}{2}\right)+\beta_0\frac{\pi^2}{4}\Bigg]\nonumber\\
                       & -\Bigg[\int\limits_1^\infty dxf(x)e^{-2xm_cr}\left(\frac53+\frac{1}{x^2}+\frac{\sqrt{x^2-1}(1+2x^2)}{2x^3}\ln\left(\frac{x-\sqrt{x^2-1}}{x+\sqrt{x^2-1}}\right)\right)\nonumber\\
                       & +\int\limits_1^\infty dxf(x)e^{-2xm_cr}\left(\ln(4x^2)-\text{Ei}(2xm_cr)-e^{4xm_cr}\text{Ei}(-2xm_cr)-\frac53\right)\nonumber\\
                       & -\left(\ln\left(\tilde{m}_cr\right)+\frac56\right)^2-\frac{\pi^2}{12}\Bigg]\nonumber\\
                       & +\frac{57}{4}\Bigg[c_1\Gamma(0,2c_2m_cr)+d_1\Gamma(0,2d_2m_cr)+\ln\left(\tilde{m}_cr\right)+\frac{161}{228}+\frac{13}{19}\zeta_3\Bigg]\Bigg\}\,,
 \label{eq:deltaV2}
\end{align}
where
\begin{equation}
 \tilde{m}_c=m_ce^{\gamma_E}, \hspace{1cm}\tilde{\mu}=\mu e^{\gamma_E}.
\end{equation}
In \eqref{eq:deltaV1} we have corrected some typos in Eq.~(30)
of~\cite{Hoang:2000fm}. This also affects an integral representation
in~\cite{Hoang:2000fm} that should read
\begin{equation}
 e^{-x}\,\text{Ei}(x)+e^{x}\,\text{Ei}(-x) =
\mathcal{P}\int\limits_0^\infty dt\frac{2te^{-x t}}{1-t^2}.
\end{equation}
However, to our understanding this does not affect other parts
of~\cite{Hoang:2000fm}.

\section{Charm effects in the relation between PS and pole mass}
\label{sec:charm_PS/pole}

The PS mass at a subtraction scale $\mu_f$ is defined by~\cite{Beneke:1998rk}
\begin{equation}
\label{eq:PS_def}
m_b = m_b^\text{PS}(\mu_f)  - \frac{1}{2} \int_{|\mathbf{q}| < \mu_f}
\frac{d^3\mathbf{q}}{(2\pi)^3} \,\tilde{V}(|\mathbf{q}|) =
m_b^\text{PS}(\mu_f) + \sum_{i=0}^\infty \delta m_i^\text{PS}\,.
\end{equation}
Defining the charm corrections to the PS-pole relation
\begin{equation}
\label{eq:delta_mc_def}
\sum_{i=1}^\infty \delta m_{i,m_c}^\text{PS} = -
\frac{1}{2} \int_{|\mathbf{q}| < \mu_f}
\frac{d^3\mathbf{q}}{(2\pi)^3} \,\tilde{V}_{m_c}(|\mathbf{q}|)\,,
\end{equation}
we obtain
\begin{align}
  \label{eq:delta_mc_1}
\delta m_{1,m_c}^\text{PS} ={}&
\frac{\alpha_s}{\pi}C_F\mu_f\frac{\alpha_s}{4\pi}\frac{4}{3}T_F
\bigg(I_\Pi(z) + I_{\ln}(m_c) + \frac{5}{3}\bigg)\,,\\
  \label{eq:delta_mc_2}
\delta m_{2,m_c}^\text{PS}
={}&\frac{\alpha_s}{\pi}C_F\mu_f\bigg(\frac{\alpha_s}{4\pi}\bigg)^2
\frac{4}{3}T_F\bigg[\notag\\
&\hspace*{-1cm}+\,\frac{4}{3}T_F\bigg(
I_{\Pi^2}(z)
 + 2I_{\Pi,\ln}(z,m_c)
 +  I_{\ln,\ln}(m_c,m_c)
 + \frac{10}{3} \big(I_\Pi(z) + I_{\ln}(m_c)\big)
 + \frac{25}{9}
\bigg)\notag\\
&\hspace*{-1cm}+\, 2\beta_0\bigg(
I_{\Pi,\ln}(z,\mu) + I_{\ln,\ln}(m_c,\mu) + \frac{5}{3} I_{\ln}(\mu)
\bigg)\notag\\
&\hspace*{-1cm}+\, 19\bigg(
c_1I_\Xi\bigg(\frac{z}{c_2^2}\bigg)
+ d_1I_\Xi\bigg(\frac{z}{d_2^2}\bigg)
+ I_{\ln}(m_c)
\bigg)\notag\\
&\hspace*{-1cm}
+\,2 a_1\bigg(I_\Pi(z) + I_{\ln}(m_c) + \frac{5}{3}\bigg)+ \frac{161}{6}
+ 26\zeta_3
\bigg]
\end{align}
with
\begin{align}
\label{eq:z}
  z ={}&\bigg(\frac{\mu_f}{2m_c}\bigg)^2\,,\\
  I_{\ln}(m) ={}& \ln\frac{m^2}{\mu_f^2} + 2\,,\\
  I_{\ln,\ln}(m_0,m_1) ={}&
  \ln\frac{m_0^2}{\mu_f^2}\ln\frac{m_1^2}{\mu_f^2} +
  2\ln\frac{m_0^2}{\mu_f^2}+ 2\ln\frac{m_1^2}{\mu_f^2} + 8\,,\\
I_\Xi(\hat{z}) ={}& 2\,\frac{\arctan\sqrt{\hat{z}}}{\sqrt{\hat{z}}} +
\ln(1+\hat{z}) - 2\,,\\
\label{eq:I_Pi}
  I_{\Pi}(z) ={}&
\frac{z}{5}\bigg[\PFQ{2}{1}{1 & 1}{\frac{7}{2}}{-z}
+ \frac{1}{3}\,\PFQ{3}{2}{1 & 1 &\frac{3}{2}}{\frac{5}{2} &
  \frac{7}{2}}{-z}\bigg]\,,\\
\label{eq:I_Pi_log}
I_{\Pi,\ln}(z,m) ={}& I_{\Pi}(z)\ln\frac{m^2}{\mu_f^2} +
\frac{2}{15}z\bigg[
\PFQ{3}{2}{1 & 1 &\frac{3}{2}}{\frac{5}{2} &\frac{7}{2}}{-z}
+\frac{1}{3}\,\PFQ{4}{3}{1 & 1 &\frac{3}{2}&\frac{3}{2}}{\frac{5}{2} &\frac{5}{2} &\frac{7}{2}}{-z}
\bigg]\,,\nonumber\\[-0.2cm]
&\\[-0.2cm]
I_{\Pi^2}(z) ={}&
\frac{1}{45\*z^3}\,\bigg\{z\*\big(-9+z\*(48+785\*z)\big)\notag\\
&\hspace*{-1cm}
-\,6\,\sqrt{z\*(1+z)}\*\big(-3+z\*(17+110\*z)\big)\*\arsinh(\sqrt{z})\notag\\
&\hspace*{-1cm}
+\,9\,(-1+5\*z+20\*z^3)\*\arsinh(\sqrt{z})^2\notag\\
&\hspace*{-1cm}
+\,z^{\frac{5}{2}}\*\Big[-18\*\pi^2+36\*\arsinh(\sqrt{z})\*\big(\ln(z)-2\*\ln(-1+\sqrt{1+z})\big)\notag\\[-0.1cm]
&\hspace*{-1cm}
\phantom{+z^{\frac{5}{2}}\*\Big(}-36\*\Li2\big((\sqrt{z}+\sqrt{1+z})^{-2}\big)+144\*\Li2\big((\sqrt{z}+\sqrt{1+z})^{-1}\big)\Big]
\bigg\}\,.\qquad\quad
\end{align}
The parameters $c_1,c_2,d_1,d_2$ are the same as
in~(\ref{eq:cd_1})--(\ref{eq:A}). $\Li2(x) = \sum_{i=1}^\infty
x^i/i^2$ denotes the dilogarithm and the ${}_PF_{P-1}$ are (generalised)
hypergeometric functions. For the sake of brevity we refrain from
rewriting the ${}_2F_1$ and ${}_3F_2$ functions in~(\ref{eq:I_Pi})
and (\ref{eq:I_Pi_log}) in terms of elementary functions. $C_F = 4/3,
C_A = 3, T_F =1/2$ are the usual group factors and $n_l=4$ is the number
of light flavours, including the charm quark. $\alpha_s$ is defined with
$n_l$ active flavours, i.e. $\alpha_s = \alpha_s^{(n_l)}(\mu)$.

Eqs.~(\ref{eq:delta_mc_1})--(\ref{eq:z}) are
parametrised in terms of the pole charm-quark mass. Employing the \MS{}
mass $m_c^{\MS}(\mu_c)$ instead, $\delta m_{2,m_c}^{\text{PS}}$ receives an
additional contribution
\begin{equation}
\label{eq:MSbar_charm_mass}
\delta m_{2,m_c}^\text{PS} \to \,\delta m_{2,m_c}^\text{PS} +
  \frac{\alpha_s}{\pi}C_F\mu_f\bigg(\frac{\alpha_s}{4\pi}\bigg)^2
  \frac{16}{3}T_F C_F\bigg(2+3\ln\frac{\mu_c}{m_c^{\MS}(\mu_c)}\bigg)
  \big(1 -z\*I_{\Pi}'(z)\big)
\end{equation}
with
\begin{equation}
I_{\Pi}'(z) =
  \frac{1}{2(z+1)}+\frac{z-2}{10(z+1)}\*\,\PFQ{2}{1}{1&1}{\frac{7}{2}}{-z}
  - \frac{1}{30}\,\PFQ{3}{2}{1&1&\frac{3}{2}}{\frac{5}{2}&\frac{7}{2}}{-z}\,.
\end{equation}
For our analysis we set $\mu_c$ to our overall renormalisation scale
$\mu$.

\section{Charm corrections to bound-state energies and
wave functions}
\label{sec:charm_psi}

Writing the bound-state energies and wave functions in the form
\begin{align}
  \label{eq:exp_bound_state}
  E_N &= E_N^{(0)} \bigg(1 + \sum_{i=1}^\infty e_N^{(i)}\bigg)\,,\\
  |\psi_N(0)|^2 &= |\psi_N^{(0)}(0)|^2
  \bigg(1 + \sum_{i=1}^\infty f_N^{(i)}\bigg)\,,
\end{align}
we can again split the higher-order terms $e_N^{(i)}, f_N^{(i)}$ into
two parts
\begin{align}
  \label{eq:split_e}
e_N^{(i)} &= e_{N,\text{massless}}^{(i)} + e_{N,m_c}^{(i)}\,,\\
  \label{eq:split_f}
f_N^{(i)} &= f_{N,\text{massless}}^{(i)} + f_{N,m_c}^{(i)}\,,
\end{align}
where $e_{N,\text{massless}}^{(i)}, f_{N,\text{massless}}^{(i)}$ are the
respective corrections for the case of massless charm quarks. A general
strategy for computing bound state corrections is described in detail
in~\cite{Beneke:2014}. We refrain from repeating this discussion
here. For $i=1$ (NLO), the mass corrections originate from a single
potential insertion into the Coulomb Green function. The $i=2$ (NNLO)
contributions can be split further into three parts
\begin{align}
  \label{eq:split_e2mc}
e_{N,m_c}^{(2)} &=  e_{N,\{m_c\}}^{(2)} +  e_{N,\{m_c,
  m_c\}}^{(2)} +  e_{N,\{m_c,
  \text{massless}\}}^{(2)}\,,\\
  \label{eq:split_f2mc}
f_{N,m_c}^{(2)} &=  f_{N,\{m_c\}}^{(2)} +  f_{N,\{m_c,
  m_c\}}^{(2)} +  f_{N,\{m_c,
  \text{massless}\}}^{(2)}\,,
\end{align}
where the subscripts $\{m_c\}$ and $\{m_c, m_c\}$ denote single and
double insertions of the
potential $V_{m_c}$, respectively. $\{m_c,
  \text{massless}\}$ denotes the contribution from the mixed double
insertion of one potential $V_{m_c}$ and one potential
$V_{\text{massless}}$. In the following we list our results for
these corrections.

\subsection{Single insertions}
\label{sec:ins_single}

The corrections to the energy levels and wave functions from single
insertions of $V_{m_c}$ read
\begin{align}
  \label{eq:e_f_charm}
  e_{N,m_c}^{(1)} &= -2\,\Gamma(1/2)\*\Gamma(N)\*\frac{\alpha_s}{4\*\pi}\*\frac{1}{2\pi i}
  \int_{-i \infty}^{i \infty}
  du\,\xi_N^{-u}\,I^{(1)}(u)
  \sum_{i=1}^N\eta_c(u,i,N)\,,\\
  e_{N,\{m_c\}}^{(2)} &= \frac{4}{3}\,\Gamma(1/2)\*\Gamma(N)\*\bigg(\frac{\alpha_s}{4\*\pi}\bigg)^2\*\frac{1}{2\pi i}
  \int_{-i \infty}^{i \infty}
  du\,\xi_N^{-u}\,I^{(2)}(u)
  \sum_{i=1}^N\eta_c(u,i,N)\,,\\
  f_{N,m_c}^{(1)} &=
-\Gamma(1/2)\*\Gamma(N+1)\*\frac{\alpha_s}{4\pi}\*\frac{1}{2\pi i}
  \int_{-i \infty}^{i \infty}
  du\,\xi_N^{-u}\,I^{(1)}(u)
  \sum_{i=1}^N\phi_c(u,i,N)\,,\\
f_{N,\{m_c\}}^{(2)} &=
\frac{2}{3}\*\Gamma(1/2)\*\Gamma(N+1)\,\bigg(\frac{\alpha_s}{4\pi}\bigg)^{\!2}
  \*\frac{1}{2\pi i}
  \int_{-i \infty}^{i \infty}
  du\,\xi_N^{-u}\,I^{(2)}(u)
  \sum_{i=1}^N\phi_c(u,i,N)\,,
\end{align}
where
\begin{align}
  \label{eq:xi_N}
  \xi_N ={}& \frac{\alpha_s C_F m_b}{2 N m_c}\,,\\
  \eta_c(u,i,N) ={}&  \frac{\Gamma(-u)\*\Gamma(u+2)}{\Gamma(N-i+1)\*\Gamma(i)^2\*\Gamma(i+1)}\frac{\Gamma(u+i)}{\Gamma(u-i+2)}\,,\\
  \phi_c(u,i,N) ={}& \eta_c(u,i,N)\,\bigg(\frac{3-u}{N}-\psi(N+1) +
  2\,\psi(i) + \psi(i+1)\notag\\[-0.2cm]
  &\phantom{\eta_c(u,i,N)\bigg(} - \psi(u+i) - \psi(u-i+2)\bigg)\,,\\
\label{eq:I1}
  I^{(1)}(u) ={}& \frac{\Gamma(2-\frac{u}{2})}{u\,\Gamma(\frac{5-u}{2})}\,,\\
  I^{(2)}(u) ={}& \frac{2\,\Gamma(3-\frac{u}{2})}
  {(u-2)\*\Gamma(\frac{7-u}{2})}-\frac{\Gamma(2-\frac{u}{2})}{u\,\Gamma(\frac{5-u}{2})}\*\Bigg(3\*a_1-3\*\beta_0\*\ln\bigg(\frac{m_c^2}{\mu^2}\bigg)\notag\\
&\quad+\,(2+3\*\beta_0)\*\bigg[\frac{2}{u}+\psi\bigg(2-\frac{u}{2}\bigg)-\psi\bigg(\frac{5-u}{2}\bigg)-\ln(
4)+\pi\*\cot\Big(\frac{\pi}{2}\*u\Big)\bigg]\Bigg)\notag\\
&+\,\bigg(\!-\frac{2}{3}\*\frac{\Gamma(\frac{u-5}{2})}{(u-6)\*\Gamma(\frac{u}{2}-2)}-2\*\frac{\Gamma(\frac{u-3}{2})}{(u-4)\*\Gamma(\frac{u}{2}-1)}+\frac{8}{3}\*\frac{\Gamma(\frac{u+1}{2})}{u\,\Gamma(\frac{u}{2}+1)}\bigg)\*\cot\Big(\frac{\pi}{2}\*u\Big)\notag\\
&-\,\frac{38}{\sqrt{\pi}}\*\frac{1}{u}\*\big(c_1\*c_2^u+d_1\*d_2^u\big)\,.
\end{align}
The integration contour is chosen such that $0 < \re(u) < 1$ on the
real axis.\footnote{The complete charm-quark contributions are actually
obtained by choosing $-1 < \re(u) < 0$. The pole at $u = 0$ then
corresponds to the contribution for massless charm quarks. Choosing a
contour with $0 < \re(u) < 1$ thus amounts to subtracting this
contribution, in agreement with our definitions~\eqref{eq:split_e},
\eqref{eq:split_f} of $e_{N,m_c}^{(i)}$ and $f_{N,m_c}^{(i)}$.}

Expressing the charm mass corrections in terms of the \MS{} mass
$m_c^{\MS}(\mu_c)$ changes the NNLO contributions as follows.
\begin{align}
  \label{eq:charm_MSbar_shift_e_f}
e_{N,\{m_c\}}^{(2)} \to{}&\,e_{N,\{m_c\}}^{(2)} +
\bigg(\frac{\alpha_s}{4\pi}\bigg)^2\bigg[\frac{4}{3} + 2\ln\Bigg(\frac{\mu_c}{m_c^{\MS}(\mu_c)}\bigg)\bigg]
\notag\\
&\times\,\bigg(- 8\Gamma(1/2)\*\Gamma(N)\*\frac{1}{2\pi i}
  \int_{-i \infty}^{i \infty}
  du\,u\,\xi_N^{-u}\,I^{(1)}(u)
  \sum_{i=1}^N\eta_c(u,i,N)\bigg)\,,\\
f_{N,\{m_c\}}^{(2)} \to{}&\,f_{N,\{m_c\}}^{(2)} +
\bigg(\frac{\alpha_s}{4\pi}\bigg)^2\bigg[\frac{4}{3} + 2\ln\Bigg(\frac{\mu_c}{m_c^{\MS}(\mu_c)}\bigg)\bigg]
\notag\\
&\times\,\bigg( - 4\Gamma(1/2)\*\Gamma(N+1)\*\frac{1}{2\pi i}
  \int_{-i \infty}^{i \infty}
  du\,u\,\xi_N^{-u}\,I^{(1)}(u)
  \sum_{i=1}^N\phi_c(u,i,N)\bigg).\qquad
\end{align}

We checked that our results for the binding energy of the
$\Upsilon(1S)$ resonance up to NNLO agree numerically
with~\cite{Hoang:2000fm}. Furthermore, the energy levels
$e_{N,m_c}^{(1)}$ as well as the wave function $f_{1,m_c}^{(1)}$ are in
numerical agreement with~\cite{Eiras:2000rh}.

\subsection{Massive double insertion}
\label{sec:ins_double_mc}

For double insertions of the potential $V_{m_c}$ we obtain the following
corrections to the binding energies and the wave functions:
\begin{align}
  \label{eq:e_double_mc}
e_{N,\{m_c,m_c\}}^{(2)} ={}&
\bigg(\frac{\alpha_s}{4\pi}\bigg)^{\!2}\frac{\pi}{\Gamma(N)^2}
\,\bigg\{\frac{h_{m_c}^{(0)}(N)}{N^2}
\big[h_{m_c}^{(0)}(N)\big(7+4N\psi(N)\big)-4h_{m_c}^{(1)}(N)\big] \nonumber\\
&+\,2\sum_{\substack{s=1\\s\neq N}}^{\infty}
\frac{h_{m_c}^{(0)}(s)^2}{s(s-N)}\bigg\}-e_{N,m_c}^{(1)} f_{N,m_c}^{(1)}\,,\\
\label{eq:f_double_mc}
f_{N,\{m_c,m_c\}}^{(2)} ={}&
\bigg(\frac{\alpha_s}{4\pi}\bigg)^{\!2}\frac{\pi}{\Gamma(N)^2}
\,\bigg\{  \frac{1}{N^2}\bigg[h_{m_c}^{(1)}(N)^2 +
h_{m_c}^{(0)}(N)\bigg(3h_{m_c}^{(0)}(N)-4h_{m_c}^{(1)}(N)[1+N\psi(N)]\notag\\
&\phantom{\quad \frac{1}{N^2}\bigg[}+h_{m_c}^{(2)}(N) + N\*h_{m_c}^{(0)}(N)\Big[4\psi(N) + N\*\big(2\zeta_2
+ 2\psi(N)^2-\psi^{(1)}(N)\big)\Big]\bigg)\bigg] \notag\\
&\quad\,- \sum_{\substack{s=1\\s\neq N}}^{\infty}
\frac{h_{m_c}^{(0)}(s)}{s(s-N)}\bigg[2h_{m_c}^{(1)}(s)-h_{m_c}^{(0)}(s)\*\bigg(2+2N\psi(N)+\frac{N}{N-s}\bigg)\bigg]\bigg\}\,.
\end{align}
The coefficients $h_{m_c}^{(j)}$ for $j=0,1,2$ are
\begin{align}
  \label{eq:h_mc}
h_{m_c}^{(j)}(s) ={}& \sum_{k=1}^{s} \frac{(-1)^k
  \Gamma(s+1)}{k\,\Gamma(k)^2\Gamma(s-k+1)} \notag\\
&\hspace*{-1cm}\times\,\frac{1}{2\pi i}\int_{-i
  \infty}^{i \infty} du\,\xi_N^{-u}\,I^{(1)}(u)\,
\frac{(k+u)\Gamma(k+u)^2\Gamma(-u)}{\Gamma(1+k+u-N)}\kappa^{(j)}(u,1+k+u-N)\,,
\end{align}
with
\begin{align}
\kappa^{(0)}(u,x) &= 1\,,\\
\kappa^{(1)}(u,x) &= u + N\psi(x)\,,\\
\kappa^{(2)}(u,x) &= u(u-1) +
N\psi(x)\big(2u+N\psi(x)\big)-N^2\psi^{(1)}(x)\,,
\end{align}
and $\xi_N$ and $I^{(1)}(u)$ as defined in
(\ref{eq:xi_N}),~(\ref{eq:I1}). We again chose a contour with $0 <
\re(u) < 1$. It should be noted that during numerical evaluation large
cancellations arise between the summands in~(\ref{eq:h_mc}) so that
typically high-precision arithmetic is required.

\subsection{Mixed double insertion}
\label{sec:ins_double_mixed}
Our result for the mixed double insertion of $V_{m_c}$ and
$V_{\text{massless}}$ is
\begin{align}
  \label{eq:e_double_mixed}
  e_{N,\{m_c,\text{massless}\}}^{(2)}
  ={}&\bigg(\frac{\alpha_s}{4\pi}\bigg)^{\!2}
  \frac{4\sqrt{\pi}}{\Gamma(N)}(-1)^{N+1}
  \bigg\{\,\frac{h_{\text{massless}}^{(0)}(N)h_{m_c}^{(0)}(N)}{N}\notag\\
  &\hspace*{-1cm}+\,\frac{h_{\text{massless}}^{(-1)}(N)}{N^2}\bigg[h_{m_c}^{(1)}(N)-h_{m_c}^{(0)}(N)\bigg(\frac{7}{2}+N\psi(N)\bigg)\bigg]\notag\\
&\hspace*{-1cm}+\,\sum_{\substack{s=1\\s\neq
      N}}^{\infty}
  \frac{h_{\text{massless}}^{(-1)}(s)h_{m_c}^{(0)}(s)}{s(N-s)}\bigg\}-e_{N,m_c}^{(1)} f_{N,\text{massless}}^{(1)}
  -e_{N,\text{massless}}^{(1)} f_{N,m_c}^{(1)}\,,\\
  \label{eq:f_double_mixed}
  f_{N,\{m_c,\text{massless}\}}^{(2)}
  ={}&\bigg(\frac{\alpha_s}{4\pi}\bigg)^{\!2}\frac{2\sqrt{\pi}}{\Gamma(N)}
  (-1)^{N+1}\bigg\{\frac{1}{2N^2}\Big[2N\Big(-Nh_{\text{massless}}^{(1)}(N)h_{m_c}^{(0)}(N)\notag\\
  &\qquad+\,h_{\text{massless}}^{(0)}(N)\big[-h_{m_c}^{(1)}(N)+h_{m_c}^{(0)}(N)\big(2+N\psi(N)\big)\big]\Big)\notag\\
  &\quad
  - \,h_{\text{massless}}^{(-1)}(N)\Big(h_{m_c}^{(2)}(N)-2h_{m_c}^{(1)}(N)\big(2+N\psi(N)\big) \notag\\
  &\qquad+\,
  h_{m_c}^{(0)}(N)\big[6+4N\psi(N)+N^2\big(\psi(N)^2-\psi^{(1)}(N)+2\zeta_2\big)\big]\Big)\Big]\notag\\
  &\quad
  +\,\sum_{\substack{s=1\\s\neq
      N}}^{\infty} \frac{1}{s(N-s)}\bigg[-h_{\text{massless}}^{(-1)}(s)h_{m_c}^{(1)}(s)-Nh_{\text{massless}}^{(0)}(s)h_{m_c}^{(0)}(s)\notag\\
  &\qquad+\,h_{\text{massless}}^{(-1)}(s)h_{m_c}^{(0)}(s)\bigg(\frac{N}{N-s}+2+N\psi(N)\bigg)\bigg]\bigg\}\,,
\end{align}
where the coefficients $h_{m_c}^{(j)}$ are as defined in~(\ref{eq:h_mc}) and
\begin{eqnarray}
  \label{eq:h-1}
  h_{\text{massless}}^{(-1)}(s) &\overset{s<N}{=}&2\beta_0 \frac{s}{N-s}\,,\\
h_{\text{massless}}^{(-1)}(N) &=& - N \big[a_1+2\beta_0\big(L_N + S_1(N)\big)\big]\,,\\
h_{\text{massless}}^{(-1)}(s) &\overset{s>N}{=}&2\beta_0 \frac{N}{s-N}\,,\\
  \label{eq:h0}
  h_{\text{massless}}^{(0)}(s) &\overset{s<N}{=}&
  \frac{1}{N-s}\Big\{-h_{\text{massless}}^{(-1)}(s)-a_1s\notag\\
&&\hspace*{-1cm}+\,2\beta_0\Big[N\Big(\psi(N-s)-\psi(N)\Big)+s\Big(S_1(N-s)-L_N\Big)\Big]\Big\}\,,\\
h_{\text{massless}}^{(0)}(N) &=& -2\beta_0\Big(1+S_1(N-1)+N\psi^{(1)}(N)\Big)\,,\\
h_{\text{massless}}^{(0)}(s) &\overset{s>N}{=}& \frac{1}{s-N}\Big\{h_{\text{massless}}^{(-1)}(s)+a_1s\notag\\
&&\hspace*{-1cm}-\,2\beta_0\Big[N\Big(\psi(s-N)-\psi(N)\Big)
+s\Big(S_1(s-N)-L_N\Big)-2\Big]\Big\}\,,\\
\label{eq:h1}
h_{\text{massless}}^{(1)}(N) &=&\beta_0\Big[\frac{1}{N}+2S_2(N-1)-N\Big(\psi^{(2)}(N)+4\zeta_3\Big)\Big]\,.
\end{eqnarray}
Here $S_i(n) = \sum_{k=1}^n k^{-i}$ denote the generalised harmonic
numbers of order $i$.

\section{NLO charm effects in the non-relativistic current correlator}
\label{sec:charm_greenfunctionl}

The perturbative expansion of $G(E)$ as defined in~(\ref{eq:Green_fun_def})
can be written as
\begin{equation}
  \label{eq:G_exp}
  G(E) = G_0(E) + \sum_{i=1}^\infty \delta_iG(E)\,.
\end{equation}
Splitting off contributions from a non-zero charm-quark mass we define
\begin{equation}
  \label{eq:delta_G_mc}
  \delta_iG(E) = \delta_iG_{\text{massless}}(E) + \delta_iG_{m_c}(E)
\end{equation}
such that $\delta_iG_{m_c}(E)$ vanishes for $m_c \to 0$.

At NLO charm mass effects enter through single insertions of the
potential~\eqref{eq:deltaV1}. Following~\cite{Beneke:2014} we split
the correction $\delta_1G_{m_c}(E)$ into a part $A$ with no additional
Coulomb exchange between the quark and the anti-quark and a part $B$
which resums all ladder diagrams with at least one Coulomb exchange. In
contrast to the cases considered in~\cite{Beneke:2014} splitting
the correction into two parts is not strictly necessary, but still helps
to elucidate the structure of the result.

Part $A$ corresponds to a two loop diagram, that will be computed in
momentum space. This diagram has no ultraviolet divergence and can
be calculated directly in $d=3$ dimensions. We obtain
\begin{align}
\delta_1G_{m_c,A}(E)  ={}&
 -\int\left[\prod\limits_{i=1}^4\frac{d^{3}\mathbf{p}_i}{(2\pi)^{3}}\right]
 \tilde{G}_0^{(0\text{ex})}(\mathbf{p}_1,\mathbf{p}_2;E)
 \delta\tilde{V}_{m_c}^{(1)}(\mathbf{p}_3-\mathbf{p}_2)
 \tilde{G}_0^{(0\text{ex})}(\mathbf{p}_3,\mathbf{p}_4;E)\notag \\
&\hspace*{-1.5cm} =\,
\frac{m_b^2\alpha_sC_F}{4\pi}\frac{\alpha_s}{4\pi}\frac{1}{2\pi i}\int\limits_{-i\infty}^{i\infty}du\,
	  \frac{u(1-u)\Gamma(-u)^2\Gamma(1/2-u)\Gamma(1/2+u)^2\Gamma(u)^2}{4\pi(\xi/2)^{2u}\Gamma(1+2u)\Gamma(5/2-u)}\,,
 \label{eq:delta1GmcA}
\end{align}
where
\begin{equation}
 \xi=\frac{\sqrt{-m_b E}}{m_c},
\end{equation}
and the contour must be chosen such that $0<\text{Re}(u)<1/2$ on the real axis.

For part $B$ we perform the calculation in position space. The result
can be written as a two-dimensional Mellin-Barnes integral
\begin{align}
\delta_1G_{m_c,B}(E)  ={}& -\int d^3\mathbf{r}
 \delta V_{m_c}^{(1)}(\mathbf{r})
 \left[G_0(0,\mathbf{r};E)^2-G_0^{(0\text{ex})}(0,\mathbf{r};E)^2\right]
 \nonumber\\
={}& \frac{m_b^2\alpha_sC_F}{4\pi}\frac{\alpha_s}{4\pi}\left(-\frac{\sqrt{\pi}}{\Gamma(-\lambda)}\right)\frac{1}{(2\pi i)^2}\int\limits_{-i\infty}^{i\infty}dw\,
            \Gamma(w+1-\lambda)\Gamma^*(-w-1)\Gamma(-w)^2\nonumber \\
         & \times\,\int\limits_{-i\infty}^{i\infty}du\, I^{(1)}(u)\xi^{-u}\frac{\Gamma(2+u)\Gamma(-u)\Gamma(1+u+w)}{\Gamma(1+u-w)},
 \label{eq:delta1GmcB}
\end{align}
with $I^{(1)}(u)$ as defined in~(\ref{eq:I1}) and
\begin{equation}
 \lambda=\frac{\alpha_sC_F}{2\sqrt{-E/m_b}}\,.
\label{deflambda}
\end{equation}

In the integral over the variable $u$
the contour should be fixed such that $0<\text{Re}(u)<1$ on the real axis.
Note that by choosing the contour to the right of the pole at $u=0$ we
have accounted for the subtraction term in the potential~\eqref{eq:deltaV1}.
The notation $\Gamma^*(-w-1)$ denotes that the contour should be chosen to
the right of the first pole at $w=-1$. For positive $E$ it can be fixed
parallel to the imaginary axis with $-1<\text{Re}(w)<0$, since the real
part of $\lambda$ vanishes and left and right poles are separated.
For positive integer values $N$ of $\lambda$ Eq.~\eqref{eq:delta1GmcB} contains
poles in $N-\lambda$ due to the pinching of the contour in the
complex $w$ plane
by left and right poles. These poles determine the charm mass corrections
to the resonances as obtained in appendix~\ref{sec:charm_psi}.

\section{Gluon condensate correction}
\label{app:nonperturbative}

The gluon condensate correction to the PNRQCD Green function
is given by~\cite{Onishchenko:2000yy}
\begin{equation}
 \delta_{\langle G^2\rangle}G(E)=
-\frac{\langle0|\pi\alpha_sG^2|0\rangle}{18}
\int d^3\mathbf{r}\int d^3\mathbf{r}'
\left(\mathbf{r}\cdot\mathbf{r}'\right)
G^{(0)}(0,\mathbf{r};E)G^{(8)}(\mathbf{r},\mathbf{r}';E)
G^{(0)}(\mathbf{r}',0;E),
\label{eq:deltaG2G_def}
\end{equation}
where the superscript $(0),(8)$ refers to the colour-singlet and colour-octet
Coulomb Green function, respectively (cf.~\cite{Beneke:2013jia}).
Proceeding as in~\cite{Beneke:2014}, we find the representation
\begin{equation}
\delta_{\langle G^2\rangle}G^{(0)}(E)=-\frac{\pi^2}{18}\,K\,
\frac{m_b^2\alpha_sC_F}{4\pi}\,\lambda^5
\sum\limits_{s=0}^\infty\frac{s!\,H_{\langle G^2\rangle}(s)^2}
{(s+3)!(s+2+\lambda/8)},
\label{eq:deltaG2G_sum}
\end{equation}
where $\lambda$ is defined in (\ref{deflambda}),
\begin{equation}
 K=\frac{\langle\frac{\alpha_s}{\pi}G^2\rangle}{m_b^4(\alpha_sC_F)^6},
\end{equation}
and
\begin{equation}
 H_{\langle G^2\rangle}(s)  = -\frac{(s+3)!}{s!}
\,\lambda\,\frac{\Gamma(5)\Gamma(s-\lambda)}{\Gamma(5+s-\lambda)}.
\end{equation}
The sum in (\ref{eq:deltaG2G_sum}) can be evaluated in terms of
polygamma functions. Expanding (\ref{eq:deltaG2G_sum}) around the bound-state
poles at $\lambda=N$ determines the condensate correction to
the S-wave energy levels $E_N$ and wave functions at the origin,
$|\psi_N(0)|^2$~\cite{Voloshin:1979uv,Leutwyler:1980tn}.

Eq.~(\ref{eq:deltaG2G_sum}) can be integrated in the complex energy plane
to yield the condensate contribution to the moments. It is worth noting
that splitting this contribution into a resonance and continuum
contribution is ill-defined, since the two are separately divergent.
For the resonances the divergence arises in the sum over $N$, since
the condensate correction rises too rapidly with principal quantum
number $N$. For the continuum, the integral over energy is too singular at
$E=0$. This reminds us that the moment calculation really refers to
the calculation of high derivatives of $\Pi_b\big(q^2\big)$ at $q^2=0$,
and assumes the validity of the operator product expansion (OPE) for
this quantity, for which the split
into resonances and continuum contributions is artificial.


\begin{table}[t]
  \centering
  \begin{tabular}{lcccccc}
    \toprule
   $n$ & 8 & 10 & 12 & 16 & 20 & 24\\
   \midrule
   $\mathcal{M}_n^{\text{exp,}\Upsilon(1S)}/\mathcal{M}_n^\text{exp}$ & 0.738 & 0.803 & 0.850 & 0.913 & 0.948 & 0.969  \\
   ${\cal \widetilde{M}}_n^{\text{pert,}\Upsilon(1S)}/{\cal \widetilde{M}}_n^\text{pert}$ & 0.769 & 0.814 & 0.849 & 0.899 & 0.932 & 0.953 \\
   ${\cal \widetilde{M}}_n^{\text{pert,rest}}/{\cal \widetilde{M}}_n^\text{pert}$ & 0.231 & 0.186 & 0.151 & 0.101 & 0.068 & 0.047  \\
   $\delta_{\langle G^2\rangle}{\cal \widetilde{M}}_n^{\Upsilon(1S)}/{\cal \widetilde{M}}_n^\text{pert}$ & 1.711 & 1.842 & 1.953 & 2.135 & 2.281 & 2.404 \\
   $\delta_{\langle G^2\rangle}{\cal \widetilde{M}}_n^{\text{rest}}/{\cal \widetilde{M}}_n^\text{pert}$ & -1.713 & -1.845 & -1.957 & -2.144 & -2.296 & -2.427 \\
   $\delta_{\langle G^2\rangle}{\cal \widetilde{M}}_n/{\cal \widetilde{M}}_n^\text{pert}$ & -0.002 & -0.003 & -0.005 & -0.009 & -0.015 & -0.023 \\
 \bottomrule
 \end{tabular}
\caption{The contribution of the $\Upsilon(1S)$ resonance and the
remaining resonances together with the continuum (rest) to the moments
is shown separately for the experimental, the perturbative, and the
gluon condensate correction to the moments. The perturbative contributions
are evaluated at NNNLO and the condensate corrections at LO with the input
values $m_b^\text{PS}(2\text{ GeV})=\mu=4.53\text{ GeV}$,
$\alpha_s=0.220486$, $m_c = 0$, and $K=0.012\text{ GeV}^4/(m_b^4(\alpha_s C_F)^6)$,
where the pole mass $m_b$ is computed from $m_b^\text{PS}(2\text{ GeV})$
at LO according to (\ref{eq:rel_pole_mass}). }
  \label{tab:Nonperturbative}
\end{table}


It {\em is} well-defined to split the condensate contribution into the one
from the $\Upsilon(1S)$ resonance and the rest. The result normalised to
the NNNLO theoretical moments (without the condensate contribution) is
given in table~\ref{tab:Nonperturbative} together with the corresponding
split-up of the perturbative contribution to the moment, and the
experimental moment.

We first note that the theoretically computed, perturbative
$\Upsilon(1S)$ contribution to the moments is very close to the
experimental one. Both are large, increasing from 80\% for the 10th
moment to more than 95\% for $n=24$. Next, we observe that the
gluon condensate correction to the $\Upsilon(1S)$ contribution
to the moment is extremely large. Taken at face value, it exceeds
the perturbative moment by a factor of about two. While there is
a large ambiguity in the absolute size of the gluon condensate
contribution (mainly related to the choice of scale in $\alpha_s$
in the expression for $K$), as discussed in~\cite{Beneke:2014qea},
the enormous cancellation between the contribution to the $\Upsilon(1S)$
resonance, $\delta_{\langle G^2\rangle}{\cal \widetilde{M}}_n^{\Upsilon(1S)}/{\cal \widetilde{M}}_n^\text{pert}$, and the remainder,
$\delta_{\langle G^2\rangle}{\cal \widetilde{M}}_n^{\text{rest}}/{\cal \widetilde{M}}_n^\text{pert}$, is independent of this size. For
$n=8$ it is effective to one part in 1000, and it remains
at the $1\%$ level even for very high moments $n=24$. As a consequence,
the total gluon condensate correction (last row in the table)
remains very small, around $2\%$, for $n=24$ when the ultrasoft
scale $m_b/n$ is clearly in the non-perturbative regime.

We believe
that the validity of quark-hadron duality must be questioned when
the experimental and perturbative contribution of the rest
is a few percent, while the condensate contribution to the same
quantity is more than 50 times larger. It appears that the gluon
condensate contribution to the entire moment is anomalously small.
Further insight into the convergence of the OPE for
high moments could be obtained from an estimate of the dimension-6
condensate contributions, which is not available. From power-counting
the breakdown of the OPE is expected when $n\Lambda_\text{QCD}/m_b\sim 1$,
which occurs in the ballpark of $n\sim 16$. If instead the maximal
value of $n$ was determined by the value at which the gluon condensate
contribution is as large as the perturbative moment, we would find
very large values of $n$, which seem to be clearly outside the range,
where quark-hadron duality can be expected to apply.


\providecommand{\href}[2]{#2}\begingroup\raggedright\endgroup

\end{document}